\documentclass[sigplan,10pt]{acmart}
\AtBeginDocument{%
  }

\renewcommand\footnotetextcopyrightpermission[1]{} 
\setcopyright{acmlicensed}
\copyrightyear{2018}
\acmYear{2018}
\acmDOI{XXXXXXX.XXXXXXX}

\acmISBN{978-1-4503-XXXX-X/2018/06}

\usepackage{circledtext}
\usepackage{amsmath,amsfonts}
\usepackage{algorithmic}
\usepackage{graphicx}
\usepackage{textcomp}
\usepackage{xcolor}
\usepackage{tikz}

\usepackage[ruled,vlined]{algorithm2e}
\usepackage{listings}
\usepackage{subfig}
\usepackage{filecontents}
\usepackage{xurl}
\usepackage{listings}
\usepackage{xspace}
\usepackage{enumitem}
\usepackage{multirow}
\usepackage{booktabs}
\usepackage{hyperref}
\usepackage{subcaption}
\usepackage{float}

\usepackage[normalem]{ulem}
\usepackage{verbatim}
\usepackage{fancyhdr}
\usepackage{pifont}

\newenvironment{ul}{
\begin{itemize}[topsep=2.5pt, partopsep=0pt, leftmargin=1em]
  \setlength{\itemsep}{2.5pt}
  \setlength{\parskip}{0pt}
  \setlength{\parsep}{0pt}
}{\end{itemize}}

\author{Shi Qiu}
\authornote{qiushijsxs@outlook.com}
\affiliation{
  \institution{Xiamen University}
  \city{Xiamen}
  \country{China}
}

\author{Yifan Hu}
\affiliation{
  \institution{Xiamen University}
  \city{Xiamen}
  \country{China}
}

\author{Xintao Wang}
\affiliation{
  \institution{Shanghai Jiao Tong University‌}
  \city{Shanghai}
  \country{China}
}

\author{Wenhao Zhu}
\affiliation{
  \institution{Xiamen University}
  \city{Xiamen}
  \country{China}
}

\author{Jianqin Yan}
\affiliation{
  \institution{Xiamen University}
  \city{Xiamen}
  \country{China}
}

\author{Hao Chen}
\affiliation{
  \institution{Xiamen University}
  \city{Xiamen}
  \country{China}
}

\author{Kaiqiang Xu}
\affiliation{
  \institution{Hong Kong University of Science and Technology}
  \city{Hong Kong}
  \country{China}
}

\author{Kai Chen}
\affiliation{
  \institution{Hong Kong University of Science and Technology}
  \city{Hong Kong}
  \country{China}
}

\author{Yiming Zhang}
\authornote{Yiming Zhang is the corresponding author. sdiris@gmail.com}
\affiliation{
  \institution{Shanghai Jiao Tong University}
  \city{Shanghai}
  \country{China}
}

\begin{document}

\newcommand{\sys}{Tutti\xspace}
\newcommand{\kv}{KV\xspace}
\newcommand{\kvs}{KV store\xspace}
\newcommand{\gs}{GStore\xspace}
\newcommand{\oo}{\textit{open()}\xspace}
\newcommand{\oor}{\textit{read()}\xspace}
\newcommand{\ow}{\textit{write()}\xspace}
\newcommand{\oc}{\textit{close()}\xspace}

\newcommand{\bluecomment}[1]{\textcolor{blue}{#1}}
\newcommand{\superscript}[1]{\ensuremath{^{#1}}}
\renewcommand{\t}[1]{\vspace{2.5pt}\noindent\textbf{#1}\xspace}
\setitemize{noitemsep,topsep=0em,parsep=0em,partopsep=3pt}
\newcommand{\qs}[1]{\textcolor{red}{#1}}
\newcommand{\hyf}[1]{\textcolor{blue}{#1}}
\newcommand{\kx}[1]{\textcolor{red}{[KX: #1]}}
\newcommand{\fact}[1]{\textcolor{red}{#1}}
\newcommand{\kvc}{KV cache\xspace}
\newcommand{\KVC}{KV Cache\xspace}

\def\todo#1{\textcolor{red}{#1}}
\newcommand*\circled[1]{\tikz[baseline=(char.base)]{
            \node[shape=circle,draw,inner sep=0.5pt] (char) {#1};}}

\hyphenation{Paged-Attention}
\pagestyle{plain} 
\title{\sys: Making SSD-Backed \KVC Practical for Long-Context LLM Serving}

\begin{abstract}

LLM serving relies on prefix caching to improve inference performance.
As growing contexts push key-value (KV) cache footprint far beyond GPU HBM and CPU DRAM capacity, 
\kvc is increasingly offloaded to NVMe SSDs.
Unfortunately, 
restoring \kvc from SSDs suffers from poor I/O performance
and incurs significant GPU stalls.
This is primarily because
the fragmented GPU memory layout results in a massive number of tiny random I/Os, 
rendering the low-parallelism CPU a severe bottleneck
even with GPU Direct Storage (GDS),
which still relies on CPU intervention to initiate each I/O
and thus remains \emph{CPU-centric}.



This paper presents \emph{\sys},
an efficient SSD-backed KV caching solution 
that eliminates CPU intervention
from the critical data and I/O control paths 
between HBM and SSDs.
At the core of \sys is a \emph{GPU-centric} \kvc object store, 
in which the CPU is only responsible for \emph{asynchronously} loading I/O kernels once per layer to the GPU.
%
\sys saturates NVMe SSD bandwidth 
and reduces GPU stalls to near zero
through the following designs:
(i) we provide a GPU-native object abstraction
that enables bulk \kvc transfers and management;
(ii) we re-architect the GPU storage stack by introducing GPU io\_uring
to support asynchronous GPU direct object I/O;
and (iii) we propose slack-aware I/O scheduling 
to avoid GPU resource contention.
We have implemented \sys and integrated it to vLLM.
Extensive evaluation shows that 
compared to the state-of-the-art GDS-enabled, SSD-backed LMCache, 
\sys reduces TTFT by 78.3\% under strict SLO constraints and improves the achievable request rate by 2$\times$. 
The serving cost is reduced by 27\%.
\sys achieves nearly the same inference performance as DRAM-backed LMCache, 
while providing almost infinite capacity.






\end{abstract}

\maketitle
\pagestyle{plain}
\section{Introduction}


%

Large Language Models (LLMs) are changing data centers 
from data storage platforms into token-generation infrastructures for AI services
\cite{multi_turn_dialogue_systems,code_understanding}. 
For Model-as-a-Service (MaaS) providers, 
the latency and cost of token generation 
determine service competitiveness. 
Prefix caching \cite{mooncake,IMPRESS,xie2025strata} has become a key optimization for modern inference serving. 
It reuses previously computed tokens, 
known as the \textit{key-value (KV) cache}, to avoid redundant computation, thereby improving Service Level Objectives (SLOs) and lowering per-token cost by up to an order of magnitude \cite{per-token-cost-deepseek,per-token-cost-openai}. 


As LLM context windows and concurrency grow, KV cache footprints rise rapidly. 
The GPU HBM is quickly exhausted, 
forcing KV eviction and recomputation that increase latency and cost while limiting the number of concurrent sessions a MaaS provider can sustain \cite{weka-kv}. 
CPU DRAM is commonly used to extend KV capacity beyond HBM, 
but still falls short at scale.
For instance,
even about 2 TB of DRAM retains only around five minutes of KV cache \cite{5min-tire}. 
Therefore, further expansion requires NVMe SSDs as the next tier 
\cite{cacheattention,cacheblend,hcache,sheng2023flexgen,mooncake,storage-next,hicache,lmcache}.
%
Commercial servers
can provide over 100 TB capacity of NVMe SSDs 
\cite{weka-kv}, 
enough to retain more than one hour of KV cache for long-running conversations and emerging agentic workloads.

However,
three-tier HBM-DRAM-SSD \kvc systems
are too slow for latency-sensitive LLM inference. 
The bottleneck is not raw SSD bandwidth \cite{D7-PS1010,KIOXIA}, 
but rather arises from the fine-grained, page-based GPU memory layout
used by modern LLM engines (vLLM \cite{vllm} and SGLang \cite{SGLang}),
which fragments a logically contiguous KV cache into many small, scattered blocks \cite{pagedattention,zheng2024sglang,TensorRT-LLM,lmcache}. 
Restoring a long prefix from SSDs generates a massive number of tiny random I/Os \cite{xie2025strata,lmcache}, 
further compounded by DRAM-HBM data copies and CPU-GPU synchronization. 
All these operations require CPU intervention,
and thus the three-tier \kvc hierarchy is \emph{CPU-centric} \cite{geminifs}.
Together, these overheads reduce effective SSD-to-GPU bandwidth 
and induce 70$\sim$80\% GPU stalls \cite{ren2025characterizing}.
%
Expensive GPU cycles are wasted 
waiting for KV cache transfers from SSDs to HBM (via DRAM), 
making KV cache reuse even slower than recomputation~\cite{flashgen,jiang2024kvpr,IMPRESS,pan2025instattention,hcache}.

A common optimization 
is to pipeline \kvc transfers with computation
to mitigate the transfer overhead,
which is effective for DRAM-backed \kvc systems \cite{hcache,flashgen}.
On SSDs, however,
pipelining would fragment transfers, 
reduce effective bandwidth, 
and introduce additional CPU-side scheduling and control overhead,
thereby further degrading I/O performance.
%
Consequently, 
existing systems tend to 
avoid offloading \kvc to SSDs
and keep most \kvc in DRAM \cite{mooncake,xie2025strata,lee2025disk,gao2024attentionstore,ye2024chunkattention,ali-kv}, whose limited capacity lowers hit rates and diminishes the benefits of prefix caching.



\begin{figure}
    \centering
     \includegraphics[width=1\linewidth]{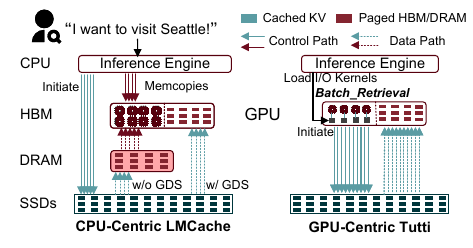}
     \caption{Comparison between CPU-centric \kvc storage (LMCache w/ and w/o GDS) and GPU-centric \sys. 
     \sys eliminates CPU intervention 
     from the critical data and I/O control paths between HBM and SSDs.
     }
    \label{fig:intro-comparision}
\end{figure}

The state-of-the-art LMCache \cite{lmcache} integrates GPU Direct Storage (GDS) \cite{GDS} to its \kvc hierarchy,
enabling an (optional) two-tier HBM-SSD mode 
with direct access between the GPU and SSDs.
However,
as each I/O must be \emph{initiated} by the CPU (Fig.~\ref{fig:intro-comparision}(\emph{left})),
GDS remains \emph{CPU-centric},
with the CPU still on the critical I/O control path. 
As a result,
GDS-enabled LMCache still suffers from I/O bottlenecks 
when transferring \kvc between HBM and SSDs
(\S\ref{sec:back-moti}).
This problem is further exacerbated as GPU compute capability and model-side efficiency continue to increase.


This paper presents \emph{\sys},
an efficient SSD-backed KV caching solution 
that eliminates CPU intervention
from the critical data and I/O control paths 
between HBM and SSDs
(Fig.~\ref{fig:intro-comparision}(\emph{right})).
At the core of \sys is a \emph{GPU-centric}, two-tier (HBM-SSD) \kvc object store, 
in which the CPU is only responsible for \emph{asynchronously} loading I/O kernels once per layer to the GPU, 
reducing CPU overhead from $O(layer \times blocks)$ to $O(layer)$.
This makes the CPU no longer a bottleneck,
enabling the GPU to issue massive parallel I/O requests for \kvc objects directly to SSDs.


Although GPU-centric storage has been explored for raw blocks (BaM \cite{bam}) and files (GeminiFS \cite{geminifs} and GoFS \cite{gofs}),
extending it to \kvc scenarios remains challenging
(\S\ref{sec:back_challenges})
due to
(i) abstraction mismatch for \kvc management,
(ii) granularity gap between KV cache transfers and GPU storage I/O,
and (iii) GPU resource contention.
\sys addresses these challenges through the following designs,
thereby
saturating NVMe SSD bandwidth
and reducing GPU stalls to near zero.

%


%


First,
we provide a GPU-native object abstraction (\S\ref{sec:design-kvstore})
that enables bulk \kvc transfers and management, 
allowing 
direct GPU access to \kvc stored on NVMe SSDs.
To achieve this,
we introduce a GPU file pool, 
an NVMe file pool 
(based on GPU file systems like GeminiFS), 
and a P2P memory mapping table.
We also expose a CPU-side interface that integrates allocation, indexing, and high-concurrency GPU access into a single operation.

%


Second, 
we re-architect the GPU storage stack (\S\ref{sec:gIO-Uring})
to support asynchronous GPU direct object I/O.
Specifically,
we introduce GPU io\_uring (gio\_uring),
which emulates the CPU-side io\_uring mechanism
to 
remove I/O submission and completion from the GPU computation critical path. 
We partition GPU resources so that I/O and compute kernels can run in parallel. 

Third,
we propose slack-aware I/O scheduling (\S\ref{sec:design-IOschedulers}) to avoid GPU resource contention for improving end-to-end inference performance.
We estimate per-layer I/O slack via offline profiling, 
and schedule \kvc transfers within these slacks
to maximize compute-I/O overlap and minimize GPU stalls.


%
We have implemented \sys and integrated it to vLLM \cite{vllm}~(\S\ref{subsec:impl}).
Extensive evaluation 
shows that 
compared to the state-of-the-art SSD-backed LMCache (with GDS), 
\sys reduces TTFT by 78.3\% under strict SLO constraints and improves the achievable request rate by 2$\times$. 
The serving cost is reduced by 27\%.
%
%

This paper makes the following contributions:

\begin{ul}
\item 
To the best of our knowledge, 
\sys is the first open-source
SSD-backed KV caching solution
that eliminates CPU intervention from the critical data and I/O control paths between HBM and SSDs.


\item We provide a GPU-native object abstraction
that bridge the granularity gap between KV cache transfers and GPU storage I/O,
together with asynchronous GPU io\_uring and slack-aware I/O scheduling.

\item We integrate \sys into vLLM,
and demonstrate its effectiveness in 
saturating NVMe SSD bandwidth and reducing GPU stalls to near zero.
SSD-backed \sys achieves nearly the same inference performance as DRAM-backed LMCache, 
while providing almost infinite capacity.
\end{ul}

%

  %

\section{Background and Motivation}\label{sec:back}
This section starts with the fundamentals of token generation and KV cache in LLM inference.
Then, we identify the inefficiency of existing tiered storage for KV cache.
Finally, we examine potential design directions to overcome these inefficiencies and highlight the key challenges in realizing such a system.

\subsection{LLM Inference and KV Cache}

\textbf{Prefill and Decode.}
Modern LLMs are built on the Transformer architecture~\cite{vaswani2017attention}.
Token generation consists of two phases: \emph{prefill} and \emph{decode}.
In the prefill phase, the model processes the input prompt in parallel, converts tokens to vectors, and computes Query ($Q$), Key ($K$), and Value ($V$) matrices.
Prefill is typically compute-bound and is measured by Time-to-First-Token (TTFT), the time to process the entire input and emit the first token.
In the decode phase, the model generates tokens autoregressively based on previously generated tokens, one step at a time.
Inter-Token Latency (ITL) is commonly used to characterize decode performance.

\t{KV Cache: Trading Memory for Compute.}
To avoid recomputing tokens during the decode phase, inference engines use a \emph{Key-Value (KV) cache} for previously computed tokens.
The $K$ and $V$ matrices produced during prefill are stored and reused for subsequent decode steps.
The KV cache is not session-bound: it can be reused across requests that share a common prompt, a technique known as \emph{prefix caching}.
When a prompt hits the cache, prefill is skipped, freeing compute capacity and reducing per-token cost by up to $\sim$90\%~\cite{per-token-cost-deepseek,per-token-cost-openai}. 
This help GPUs to generate tokens faster and sustain higher QPS, improving SLOs and user experience.

\t{Paged KV Memory Management.}
\kvc footprint grows with input length, and variable-sized requests cause fragmentation in GPU memory.
To address this problem, modern inference systems~\cite{pagedattention} partition the \kvc into non-contiguous blocks of shape $[{\text{Block}}, h, d]$ along the layer and token dimensions, where each block usually holds 16–32 tokens.
Blocks are allocated on demand to support dynamic sequence growth and align with layer-wise computation.
This paged layout has become the de facto standard in modern LLM inference engines such as vLLM~\cite{vllm}, SGLang~\cite{SGLang}, and TensorRT-LLM~\cite{TensorRT-LLM}.


\begin{figure}
    \centering
    \includegraphics[width=1\linewidth]{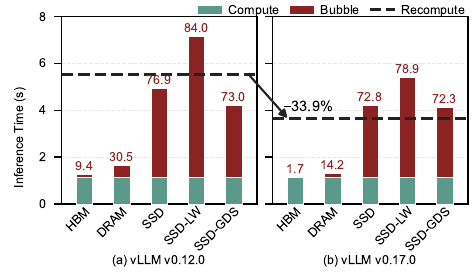}
    \caption{Inference performance of vLLM with LMCache on Llama3-8B, across HBM, DRAM, and SSD tiers (sequenth length = 64K, hit rate = 75\%). DRAM remains close to HBM, whereas SSD and GDS incurs large GPU bubbles. 
    The dashed line marks recomputation performance.
    As LLM engines continuously optimize inference computation,
    restoring \kvc from SSDs is no longer beneficial (vLLM v0.12.0 vs. v0.17.0)
    due to severe I/O bottleneck.
    %
    %
    }
    \label{fig:motivation-exp-bubble}
\end{figure}


\subsection{SSD-Induced Bottlenecks in Tiered KV Cache}\label{sec:back-moti}

As context windows scale to millions of tokens~\cite{gemini_1.5,llama4} and the number of active sessions grows, the aggregate \kvc footprint quickly exceeds GPU HBM capacity~\cite{storage-next}.
\kvc offloading extends GPU HBM capacity with CPU DRAM and NVMe SSDs,
resulting in the two-tier HBM-DRAM and three-tier HBM-DRAM-SSD hierarchies.
The HBM-DRAM hierarchy only incurs slight performance degradation,
but the extended capacity is limited.
In contrast,
the HBM-DRAM-SSD hierarchy provides much higher capacity, 
but causes significant I/O overhead.
%
%

When offloading \kvc to SSDs,
the main challenge stems from the mismatch between paged KV layouts and SSD access patterns.
Once non contiguous GPU KV blocks are evicted to the SSD, memory fragmentation~\cite{shen2024fastswitch} becomes severe I/O fragmentation.
For a 64-layer Qwen3-32B model~\cite{qwen3} with block size 64, reloading a 128K-token \kv requires fetching about 256~K (= $2 \times 64 \times 128 \times 1024 / 64$) scattered 80~KB objects.
This access pattern generates a massive number of small, random transfers, causing CPU-GPU copy, file system, and I/O submission overheads~\cite{shen2024fastswitch,xie2025strata,bam,geminifs,gofs} to dominate data movement.
Grouping multiple blocks into larger chunks can improve I/O efficiency, but introduces a tradeoff among transfer efficiency, prefix-sharing effectiveness, and cache-management granularity.
For example, the default LMCache~\cite{lmcache} chunk stores 256 tokens, causing a 128K-token \kv to require more than 1,000 chunk accesses, most of which are random.
With compute-I/O pipelining, the number of accesses further grows to tens of thousands.
As a result, 
expensive GPU cycles are wasted waiting for restoring \kvc from SSDs,
making \kvc reuse even slower than recomputation.

\t{SSD Tiers Cause Growing GPU Bubbles.}
To examine how these bottlenecks manifest in practice, we use the latest version (v0.4.2) of LMCache \cite{cacheattention,lmcache} as the representative tiered \kvc store.
LMCache supports DRAM and SSD tiers, layer-wise compute-I/O pipelining~\cite{hcache,flashgen}, and (optional) GPU Direct Storage (GDS)~\cite{GDS}.
%
%
We evaluate Llama3-8B on different vLLM versions 
(v0.12.0 released on Dec. 2025 vs. v0.17.0 on Mar. 2026) with a 64K sequence length at 75\% hit rate, with 50~GB/s DRAM-HBM bandwidth and two SSDs with peak bandwidth of 29~GB/s for read and 12~GB/s for write
(See \S\ref{sec:eval} for detailed configurations).  

\circled{1}~\textit{DRAM tier is efficient.}
As shown in Fig.~\ref{fig:motivation-exp-bubble}, loading KV from CPU DRAM introduces only modest overhead relative to HBM.
Low-latency, fine-grained DRAM-HBM access, together with LMCache's GPU-assisted copy, collapses many sequential \texttt{cudaMemcpyAsync} calls into a small number of GPU kernels with minimal control overhead.
In addition, DRAM's low latency and strong random-access performance allow layer-wise pipelining to effectively hide data movement behind attention computation.
%

%

\circled{2}~\textit{SSD tier is inefficient even with GDS.}
When extending the hierarchy to SSDs, restoring \kvc becomes highly inefficient
even with aggregated KV transfer and asynchronous I/O~\cite{didona2022understandingio}.
As shown in Fig.~\ref{fig:motivation-exp-bubble}, 
restoring \kvc from SSDs performs much worse than from DRAM, 
causing GPU bubbles to exceed 70\% of total inference latency in all cases.
%
Applying layer-wise transfers on SSDs (SSD-LW) further reduces I/O granularity and increases the number of operations, inflating end-to-end latency and pushing GPU bubble time to around 80\% of total inference latency.

GDS~\cite{GDS} removes CPU-GPU copies through peer-to-peer DMA, 
but still relies on CPU intervention to initiate each I/O,
incurring substantial software overhead 
and limiting I/O parallelism~\cite{DeepNVMe,gofs}.
Even with GDS, GPU bubble time remains high at above 70\%, indicating that 
eliminating the CPU from the data path alone 
hardly alleviates the mismatch between paged KV layouts and SSD
access patterns.
Moreover,
as LLM engines continuously optimize inference computation,
restoring KV cache from SSDs is no longer beneficial due to severe I/O bottleneck.

%
%

\subsection{GPU-Centric Storage}\label{sec:back_kvstore}
GPU-centric storage~\cite{smartio,geminifs,bam,gmt} moves both the data plane and the I/O control plane onto the GPU.
It enables GPU threads to issue NVMe I/O without CPU intervention.
BaM~\cite{bam} was the first to manage NVMe Submission Queues (SQ) and Completion Queues (CQ) directly in GPU memory,
so that GPU kernels can enqueue I/O commands, ring the NVMe doorbell, and observe completions entirely from device code.
This reduces CPU-GPU synchronization and kernel launch overhead, allowing massively parallel GPU threads to drive high-bandwidth, fine-grained I/O.


\t{Common GPU-centric storage abstraction.}
Across GPU-centric storage systems, GPU threads interact with a high-throughput software cache (e.g., an array in BaM or a page cache in GeminiFS~\cite{geminifs}) through a block or file interface.
On a cache miss, a GPU thread enqueues an I/O request into the NVMe submission queue in GPU address space and rings the doorbell register.
It then polls the completion queue until data arrives.
By staggering the I/O and compute phases of different warps, this GPU-centric design can overlap computation and storage access and hide latency.

\t{Implications for \kvc workloads.}
While GPU-centric storage provides a promising direction—GPU-controlled, fine-grained access to NVMe—it is designed around generic block and file abstractions and keeps busy-waiting at the thread or warp level.
As we show next, this abstraction does not align well with the KV cache layout and tightly pipelined decode in LLM inference, leading to problems including excessive control overhead, poor request coalescing, and underutilized NVMe bandwidth when applied naively to tiered KV storage.

\vspace{-1mm}
\subsection{Challenges of GPU-centric Storage for KV Cache}\label{sec:back_challenges}


Applying GPU-centric storage to \kvc workloads faces unique challenges in abstraction, granularity, and contention.

\begin{figure}[t]
    \centering
    \includegraphics[width=1\linewidth]{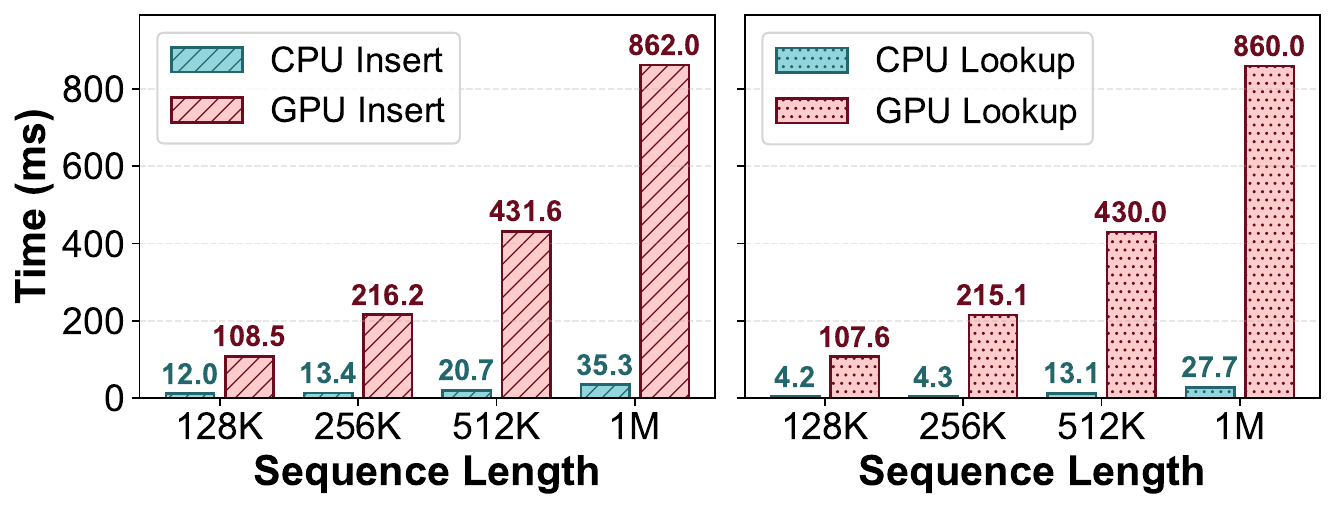}
    \caption{CPU vs. GPU in Hash Performance.}
    \label{fig:metadata-pushdwon}  
\end{figure}

\t{Abstraction mismatch for KV cache management.}
LLM engines (vLLM \cite{vllm} and SGLang \cite{SGLang}) need dynamic GPU memory block allocation and indexing for \kvc, while GPU-centric storage exposes only low-level disk block and file interfaces.
Pushing this management down to the GPU requires implementing hash based allocation and lookup in device code.
However, as shown in Fig.~\ref{fig:metadata-pushdwon}, GPU hash tables perform poorly: with various sequence lengths, insert and lookup costs are higher than CPU hash tables by $9.0\times\sim24.2\times$ and $25.6\times\sim50.0\times$, 
respectively,
up to seconds per operation.
This is hard to fix because hash computation and probing form a sequential dependency chain, and each block’s hash depends on the previous one.
Such irregular, pointer chasing workloads map poorly to SIMT execution and cannot exploit GPU parallelism. 

\t{Granularity gap between    storage I/O and KV transfers.}
The GPU NVMe driver is optimized for fine-grained, cache-like access, but KV cache reloads require medium-size, contiguous transfers to meet SSD bandwidth targets.
On PCIe 5.0 SSDs~\cite{Solidigm,KIOXIA}, 4KB requests can saturate IOPS, yet only use about 80\% of read bandwidth and 16\% of write bandwidth, 
resulting in significant underutilization of available throughput.
Simply increasing request size is nontrivial.
The GPU NVMe driver relies on NVMe Physical Region Pages (PRPs) to describe GPU HBM addresses to the controller.
Fixed 4~KB PRPs can be pre-allocated by the CPU driver, but \kvc transfers are variable and much larger ($\sim$100~KB).
For requests above 8~KB, NVMe needs additional PRP list pages, whose allocation and address translation must be done in privileged CPU code~\cite{nvme2,geminifs}.
Because GPU programs run unprivileged, GPU-centric storage cannot easily coarsen I/O without falling back to the CPU, 
which undermines the goal of eliminating CPU intervention.

\t{Resource contention.}
GPU-centric storage I/O competes with LLM computation for resources.

\circled{1} \emph{SM competition.}
LLM inference has strict data dependencies: attention cannot proceed until the corresponding \kvc is available.
Existing GPU-centric storage designs perform synchronous, busy-waiting I/O, where GPU threads continuously poll completion queues inside the compute kernel and block computation.
Without careful decoupling, 
simply adding more I/O parallelism can only reduce the SM budget available for computation. 

\circled{2} \emph{Bandwidth competition.}
Prefix caching generates heavy, bidirectional traffic.
Particularly during compute-I/O pipelining, simultaneous writes (from the previous layer) and reads (for the next layer) cause contention for NVMe resources (e.g., SSD internal cache), degrading NVMe bandwidth (\S\ref{sec:design-IOschedulers}).
Achieving fine-grained I/O orchestration within the GPU to improve utilization is difficult, while separate scheduling tends to increase kernel execution time.

 \section{Design and Implementation}

%


%


In this section, we introduce the design and implementation of \textit{\sys}.
We first describe the GPU-native object abstraction that enables high-concurrency GPU direct access to \kvc using object semantics.
We then explain how applications can efficiently submit and reap asynchronous GPU I/O kernels. 
Finally, we discuss how to schedule GPU I/O kernels to 
minimize resource contention.
%


\subsection{GPU-Centric Object Store} \label{sec:design-kvstore}
At the core of \sys is a GPU-centric \kvc object store. 
%
As discussed in \S\ref{sec:back_challenges},
\kvc management cannot be pushed entirely onto the GPU: indexing, global sharing across requests, and engine-visible mapping must remain coordinated with the CPU-side inference engine.

Fortunately,
all the management logic can be handled by the CPU \emph{off} the critical data and I/O control paths of \kvc transfers. 
We therefore build our GPU-centric object store
upon GeminiFS~\cite{geminifs},
a \emph{companion} file system for GPUs
which coexists with a conventional CPU-side file system (like ext4)
so that
the file system metadata can be managed on the CPU and shared with the GPU.
%
We extend GeminiFS with a scalable GPU file pool and a P2P memory mapping table for dynamic, bulk KV cache transfers and KV management operations such as \texttt{Store} and \texttt{Retrieve}).

\begin{figure}[t]
    \centering
    \includegraphics[width=0.8\linewidth]{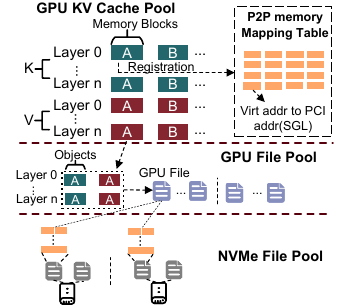}
    \caption{Layout of GPU-centric \kvc store.}
    \label{fig:data layout}
\end{figure}

\t{Scalable GPU File Pool.}
As shown in Fig.~\ref{fig:data layout}, \sys aligns storage allocation with the inference engine's KV block manager, 
by representing each memory block as one object.
A GPU file is organized as $2 \times L$ objects ($L$ is the number of layers), one key object and one value object for each layer.
This mapping preserves the inference engine's native block granularity, making dynamic allocation, indexing, and sharing consistent across HBM and SSD tiers.

GPU files are visible to the inference engine, 
while NVMe files are managed by GeminiFS
as physical storage extents allocated to SSDs.
\sys maps each GPU file to multiple NVMe files using the \textit{Tensor-Stripe} layout, which follows the original tensor granularity instead of fine-grained storage striping.
Consequently, the GPU file shape matches the \kvc memory object ($2 \times layer \times block$), so storage I/O remains aligned with KV transfer granularity.
%
%
For prompts spanning multiple GPU files, we employ a round-robin placement strategy across devices. Specifically, objects are uniformly distributed across multiple NVMe SSDs in a row-sequential manner. This approach not only balances I/O traffic across the drives to help saturate aggregate NVMe bandwidth but also reduces the indexing overhead between GPU and NVMe files.

At system startup, \sys pre-allocates a large pool of NVMe files on each device and exposes them as free GPU files.
When a new \kvc needs to be persisted, the runtime only selects an empty GPU file and installs a CPU-side hash mapping from the \kvc to the GPU file ID.
This preserves the dynamic allocation semantics expected by the runtime while removing file creation, reclamation, and other metadata operations from the runtime critical path.
%
  
%


%
\t{P2P Memory Mapping Table.} %
The GPU file pool solves logical object management, while the remaining challenge is to translate \kvc virtual addresses into PCI-visible physical addresses during runtime I/O submission.
Because modern inference engines pre-allocate a fixed \kvc memory pool at initialization and keep it stable throughout the process lifetime, \sys can pre-compute a P2P memory mapping table at startup and reuses it for subsequent GPU I/O.


However, a straightforward PRP-based design causes significant memory overhead.
For instance,
for a 60 GB KV cache on 80 GB HBM, PRP requires a pointer for every page ($\text{Total Pages} = 60 \times 1024^3 / 4096 = 15,728,640 \text{ Pages}$).
If allocating PRP List Pages at 64KB granularity (where each page holds only 16 pointers),
$983,040$ pages are required. 
This results in an actual HBM usage of ($983,040 \times 4 \text{ KB} \approx) 3.75 \text{ GB}$, significantly wasting the expensive HBM resource.




To better match medium-sized KV transfers, \sys adopts Scatter Gather Lists (SGL)~\cite{nvme2} rather than PRP.
It uses only 16 Bytes to describe a large chunk of contiguous memory, containing a Physical Address (8 bytes), Length (4 bytes), and Identifier (4 bytes). 
%
%
Consequently, memory consumption drops to ($983,040 \times 16 \text{ B} \approx) 15 \text{ MB}$.

%

At runtime, the inference engine only performs block lookup and P2P table lookup to generate a batch of lightweight \emph{GPU I/O contexts}, which are then passed to the GPU for concurrent execution.
This avoids per-request physical address construction and file-management overhead on the critical I/O path.
%
Engine-visible mappings remain CPU-managed, while the GPU holds only the metadata required for direct I/O submission.
Thus, \sys provides layer-wise batched \texttt{Store} and \texttt{Retrieval} interfaces that reduce CPU overhead from $O(layer \times blocks)$ to $O(layer)$. 

\begin{figure}[t]
    \centering
    \includegraphics[width=1\linewidth]{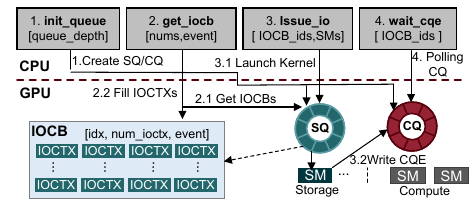}
    \caption{Architecture and I/O Process of GPU io\_uring.}
    \label{fig:desigen-giouring}
    \vspace{-1mm}
\end{figure}

\subsection{GPU io\_uring} \label{sec:gIO-Uring}
The GPU-centric object store follows a ``CPU-prepared, GPU-executed'' model.
The CPU runtime prepares I/O control blocks (IOCBs) from CPU-managed mappings and enqueues GPU I/O kernels ahead of time together with model-compute kernels.
Once enqueued, GPU-side dependency tracking determines when SSD access is issued, so the runtime I/O critical path no longer involves the CPU.
This naturally decouples GPU I/O from the computation kernel while enabling efficient parallelism between the I/O kernel and the computation kernel.
Since its design mirrors the CPU-side io\_uring~\cite{didona2022understandingio}, we call it \textit{GPU io\_uring} (gio\_uring), the architecture of which is shown in Fig.~\ref{fig:desigen-giouring}.


\t{Zero-Copy Ring Buffers.}
To avoid runtime memory allocation, copying, and CPU-GPU synchronization when the CPU prepares GPU I/O work ahead of execution,
%
gio\_uring utilizes a pair of lock-free ring buffers (SQ and CQ) residing in GPU HBM but mapped to the CPU via non-cached mmap~\cite{yan2025phoenix}.
To accommodate thousands of concurrent GPU I/O requests, the system uses a batching queue structure.
In contrast to the traditional CPU io\_uring (where one SQ entry corresponds to a single command), each SQ entry is defined as an I/O IOCB, with each IOCB containing 2048 I/O contexts (IOCTXs).

An IOCTX records the SGL address, GPU file offset, and length. 
The number of IOCTXs aligns with the GPU's minimum scheduling unit. 
For example, on an H100 (where the unit is 2 SMs), each SM supports 64 Warps of 32 threads, totaling 4096 concurrent threads. 
Considering register pressure, we typically divide the theoretical limit by 2. This design allows GPU submit massive I/O requests at once. 
%

\t{SM Partitioning For Accurate I/O.}
Simple concurrency using multiple CUDA streams is insufficient
for achieving fine-grained overlap between computation and I/O.
Due to the largely non-preemptive nature of the GPU's hardware scheduler~\cite{lin2025bullet}, a long-running I/O kernel can monopolize resources and block the execution of a critical compute kernel on another, even if idle SMs are available.
Through NVIDIA green context~\cite{green-contex}, we isolate GPU resources at the hardware level into a ``Compute Domain'' and an ``I/O Control Domain''. 
The I/O control kernel runs on dedicated SMs, unaffected by compute workload fluctuations. This ensures that latency-sensitive kernels start and complete as quickly as possible.
This design avoids long-tail latency and resource starvation 
which are common in traditional cooperative multitasking, 
and provides deterministic QoS.


\t{Async I/O Processing}: The processing of gio\_ring is similar to the conventional CPU-side io\_uring:
\circled{1} \textit{init\_queue(depth)} creates an SQ and CQ containing depth IOCBs, each with a unique index.
\circled{2} \textit{get\_iocb(nums, event)} is called before execution to retrieve the necessary IOCBs. The application fills them with CPU-side virtual addresses and updates \textit{num\_ioctx}. To maintain correctness under out-of-order stream execution, a CUDA event is inserted so that the GPU I/O kernel starts only after the required dependency is satisfied.
\circled{3} \textit{issue\_io(IOCB\_ids, SMs)} enqueues a GPU I/O kernel with the specified IOCB IDs and SM allocation, realizing intra-device parallelism. After the kernel is enqueued, SSD commands are generated and issued entirely on the GPU. When the kernel completes, it atomically writes the IOCB index to the CQ.
\circled{4} \textit{wait\_cqe()} provides fine-grained waiting by checking the CQ for a specific IOCB index without requiring CPU participation in per-I/O issuance.





%


\begin{figure}[t]
    \centering
    \includegraphics[width=1\linewidth]{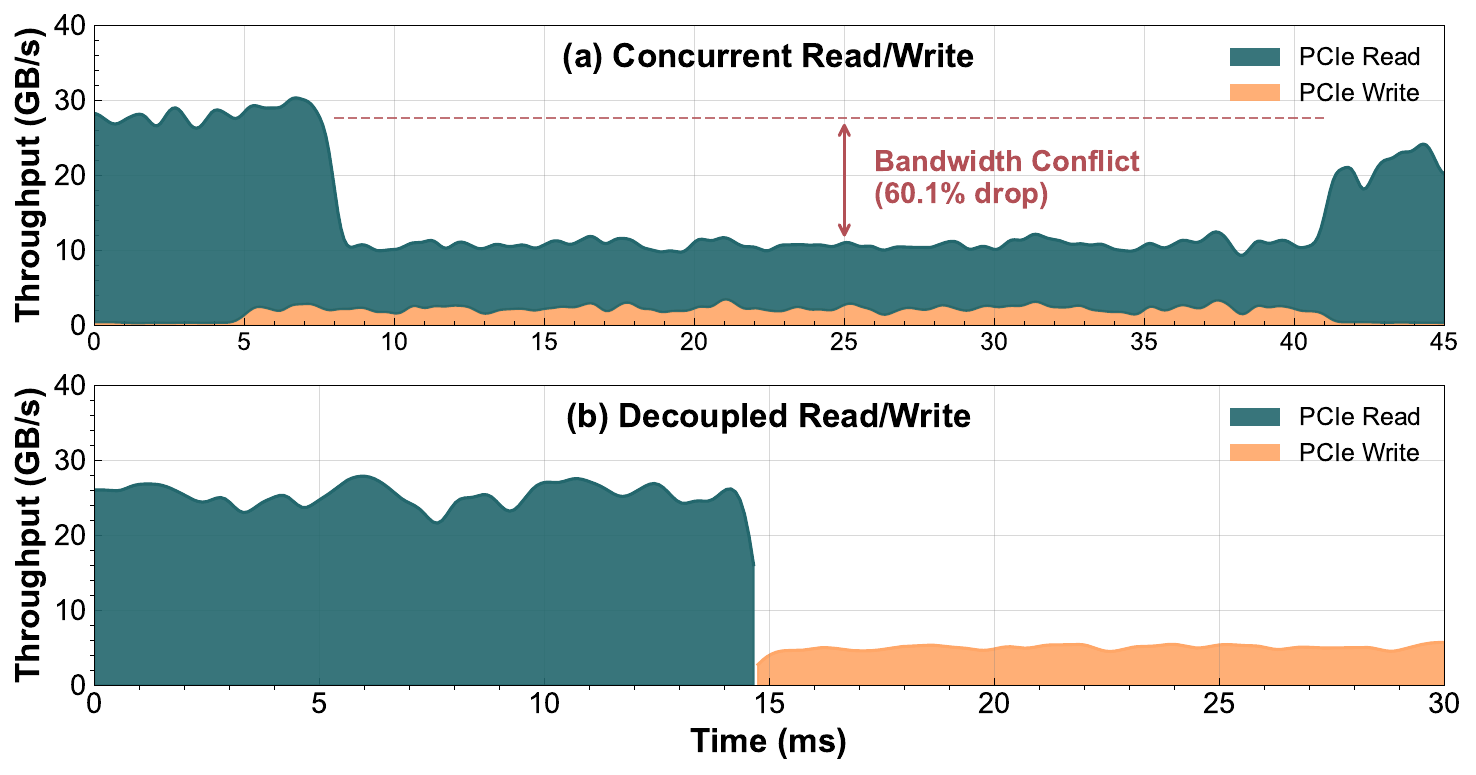}
    \caption{Concurrent vs. decoupled read/write PCIe bandwidth utilization.}
    \label{fig:Concurrent}
\end{figure}

\subsection{Slack-Aware I/O Scheduler} \label{sec:design-IOschedulers}
Simply using asynchronous I/O and SM partitioning is insufficient for achieving stable compute-I/O overlap,
as two sources of interference remain.
First, read and write I/O contend for SSD bandwidth and internal resources.
This is common in naive layer-wise pipelining, where write I/O for newly generated KV competes with read I/O for the next layer.
The loss is not a simple additive sharing effect: as shown in Fig.~\ref{fig:Concurrent}, total bandwidth drops by 60\% under concurrent read/write, whereas separate calls can saturate the device.
This is mainly because large-block reads and writes contend for the NVMe's internal cache~\cite{liu2022improving,hu2015pass}, and we reproduced this behavior with FIO using one read thread and one write thread at 256 MB granularity.

Second, I/O kernels also compete with model execution for SM resources.
Operators such as embedding, normalization, and GEMM may require up to 90\% of GPU resources.
Under non-preemptive GPU scheduling, a long-running I/O kernel can therefore delay critical compute kernels and reduce inference performance.

\begin{figure}[t]
    \centering
    \includegraphics[width=1\linewidth]{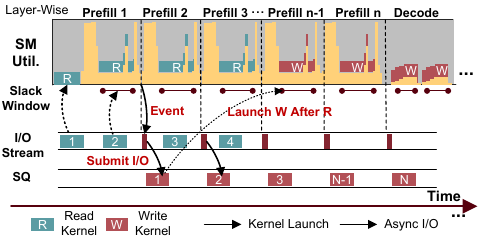}
    \caption{Slack aware I/O scheduler.}
    \label{fig:design-scheduler}
\end{figure}

To address both effects, \sys proposes a lookup-table-driven slack-aware I/O scheduler, as shown in Fig.~\ref{fig:design-scheduler}.
Slacks refer to execution windows with spare SM resources and without harmful read/write bandwidth contention.
The scheduler uses offline profiles to place read and write kernels only into such windows, thereby minimizing interference with model execution.

\t{Offline Profiling of SM Slack Windows.}
Prefill complexity varies with prefix length.
The primary source of variation is attention complexity.
As prefix length increases, the number of attention operations per new token increases linearly, leading to higher FLOPs compared to the zero-prefix baseline.
Conversely, other operators within a layer (such as Linear Projections and Normalization) are unaffected by context length.
Therefore, we profile each layer offline and store the resulting slack information in a lookup table indexed by input length ($L_{input}$) and prefix length ($L_{prefix}$).
Each entry records the duration and available SM budget of schedulable slack windows, allowing the runtime to directly look up how many IOCBs can be launched without online modeling.
The step size aligns with the token length of a single warp, drastically reducing both the offline profiling time and data size.
Additionally, we profile decode duration and the execution time and SM occupancy of read/write kernels under different IOCB counts, enabling the scheduler to select an appropriate launch size by table lookup.

%
%

%








%

\t{Decoupled Scheduling for Read and Write.}
%
To avoid the bandwidth collapse caused by concurrent read/write execution, \sys does not use naive layer-wise pipelining that overlaps reads and writes indiscriminately.
Instead, it schedules them separately according to the profiled slack table.
During prefill, read kernels have higher priority because KV retrieval lies on the critical path of reuse.
When an inference request arrives, the runtime enqueues the corresponding read IOCBs.
Before each layer begins, the scheduler consults the lookup table using the current input length and prefix length, then launches the maximum IOCB count that fits within the next profiled slack window.
If no suitable slack window exists, high reuse has made KV retrieval the bottleneck, and the scheduler immediately launches the required reads to avoid stalling computation.

Write requests are handled only after the critical-path reads have been scheduled.
Pending writes remain recorded in SQ, and gio\_uring automatically inserts CUDA events to preserve correctness.
If the current prefill layer still exposes a schedulable slack window, the scheduler issues as many writes as the lookup table allows; otherwise, it defers them to shorten prefill and preserve TTFT.
Remaining writes are flushed during decode using a best-effort policy.
Although decode usually offers lower GPU utilization, its slack windows are short and less predictable, so the scheduler relies on table lookup to opportunistically issue writes.
Requests that do not fit remain queued for later slack windows, reducing inter-request interference and improving throughput.

\subsection{\sys Implementation}\label{subsec:impl}

\t{Integration with vLLM.}
We implemented \sys using $\sim$8,000 LoC in C++ and integrated it with vLLM's KVConnector in multiple versions using $\sim$1,500 LoC in Python.
This integration preserves vLLM's block-granular KV management.
The GPU file pool exposes layer-wise \texttt{retrieve\_layer} and \texttt{store\_layer} interfaces to support efficient layer-wise KV movement in vLLM.
This organization matches the layer-wise transfer model described earlier and creates opportunities to overlap KV movement with model computation.
 The extension to vLLM is used to register the pre-allocated KV memory block pool, identify reusable prefixes, 
 and construct the mapping from logical KV blocks to GPU files.

Retrieve\_layer is issued on the critical path of reuse, while store\_layer is queued and deferred when necessary so it can be flushed in later slack windows, including subsequent requests, thereby reducing inter-request interference.
To preserve correctness and limited GPU resource usage, these interfaces are bound to CUDA stream dependencies, while the detailed GPU-side submission and completion flow follows Sec.~\ref{sec:gIO-Uring}.
Scheduling decisions are then delegated to the slack-aware scheduler.

During the warm-up, \sys profiles the per-layer slack windows for a given model and system configuration.
The resulting profile only needs to be generated once and can be reused across inference processes under the same deployment setting.
Before each retrieve\_layer or store\_layer call, the runtime consults the current layer's slack entry to decide whether to issue I/O and how many IOCBs to launch, thereby minimizing contention with inference kernels.

\t{Support for Multi-GPUs.}
When the model uses multi-GPU deployment such as tensor parallelism, vLLM launches one process per GPU, allowing one \sys instance to be deployed alongside each GPU process.
Each \sys only manages part of \kvc, each process are independently responsible for the KV blocks corresponding to its GPU-resident layers, and the size of GPU file will adjust accordingly.
To support NVMe sharing between GPUs, we use a local daemon that allocates GPU memory and initializes a dedicated NVMe submission/completion queue pair for each GPU.
The corresponding vLLM process obtains the addresses of its GPU-resident queues through GPU inter-process shared memory and submits I/O commands directly through them.
Because each GPU owns an independent queue pair, there is no inter-GPU queue contention, allowing all GPUs to access local NVMe in parallel for high-throughput KV reads and writes.
The Solidigm D7-PS1010~\cite{D7-PS1010} used in our prototype support up to 256 I/O queues, allowing us to provision 32 queues for each of 8 GPUs. This queue count is already sufficient for \sys to fully utilize the bandwidth of a single SSD.

\t{Scalability.}
To scale beyond a single node, \sys combines its local high-performance storage data plane with a distributed coordination layer.
In this design, \sys remains the per-server fast path for GPU-to-local-NVMe KV transfers, while Mooncake~\cite{mooncake} serves as the cluster-wide control plane for space allocation, replica metadata management, and location lookup.
This separation preserves the low-latency local path of \sys while allowing KV cache capacity and reuse to scale across inference servers.

When KV cache is evicted from GPU memory, the inference engine first requests space allocation from Mooncake.
\sys then persists the KV tensors to local NVMe SSDs through its P2P DMA path.
After the write completes, it notifies Mooncake to register the resulting replica metadata, making the offloaded KV globally discoverable for future reuse.

When a request needs to reuse a historical KV cache, the runtime first queries Mooncake for the candidate replica locations.
The system follows a local-first routing policy.
If a local replica is available, \sys directly loads it into GPU memory through \sys.
Otherwise, the request falls back to a remote retrieval path, where the data is fetched from a remote node and then delivered to the local GPU.

Our current prototype does not yet optimize this remote path.
It uses a CPU-side interface to read the GPU file into host memory and then transfers it across nodes via RDMA, which minimizes changes to Mooncake but adds extra CPU overhead.
In future work, we plan to extend the design to support a more direct GPU-driven remote path, for example by staging data in GPU memory and then issuing GPU-initiated RDMA to the destination GPU.


 \section{Evaluation}\label{sec:eval}
\begin{figure*} 
    \centering
    \includegraphics[width=1.0\linewidth]{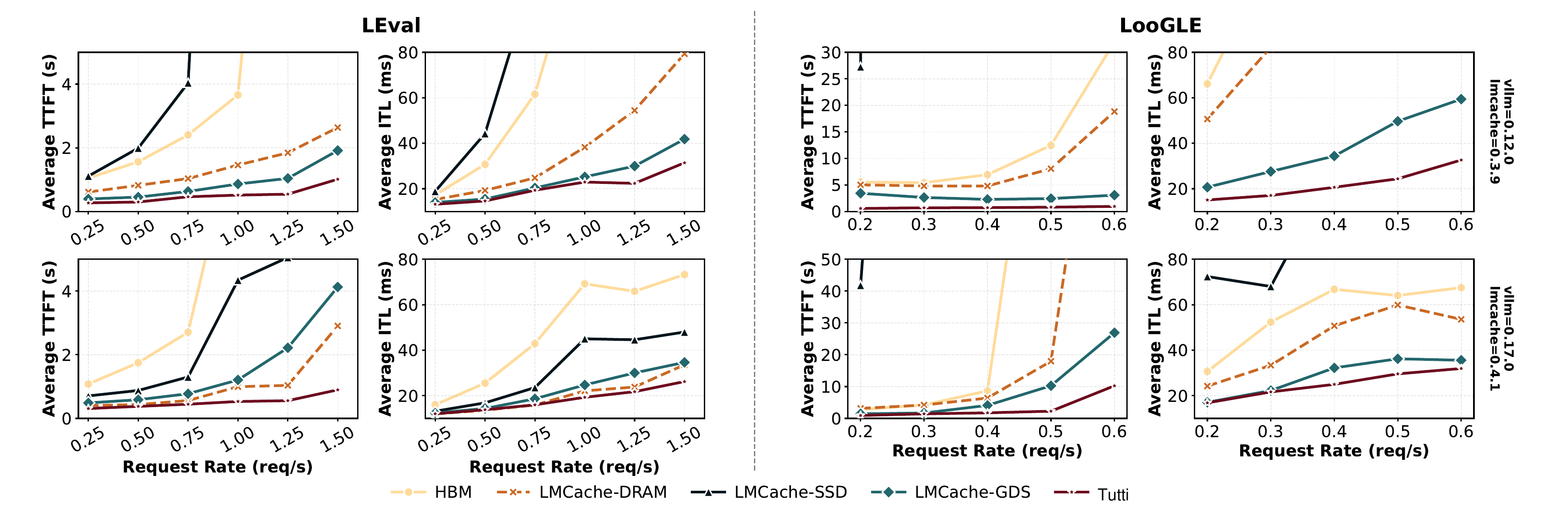}
    \caption{End-to-end TTFT and ITL on Llama3-8B across LEval and LooGLE under two vLLM versions (v0.12.0 vs. v0.17.0) with the latest LMCache. 
    As request rate increases, \sys maintains the lowest and most stable latency curves, consistent with the end-to-end analysis that its storage-compute co-design remains effective across versions. 
    Data points are omitted when systems violate SLO constraints. %
    }
    \label{fig:e2e}
\end{figure*}
We have conducted a series of experiments to evaluate the effectiveness of \sys, focusing on the following two critical questions:
\begin{enumerate}[leftmargin=*]
    \item How does \sys perform in terms of end-to-end latency for LLM inference      compared to state-of-the-art $\text{KV}$ cache services?
    \item How do the components of \sys contribute to and optimize the final inference latency and overall system efficiency?
\end{enumerate}
\noindent\textbf{Environments.}
We deployed \sys on a 64-core Intel Xeon 6530 server equipped with 512 GB of memory.
The server is equipped with two H100 GPUs with 80GB HBM and $4\times$ Solidigm D7-PS1010 7.68TB enterprise SSDs~\cite{D7-PS1010}.
%
For tiered-storage configurations, we allocate 256 GB host DRAM as pinned memory and provision 14 TB of SSD volume for each GPU.


\noindent\textbf{Baselines.}
We compare \sys against baselines from two generations of vLLM: vLLM 0.12.0 and vLLM 0.17.0.
This setup allows us to examine how improvements in serving-side compute efficiency affect end-to-end system behavior.
LMCache optimizes data movement by aggregating tokens into coarse-grained chunks (e.g., 256 tokens) to maximize SSD bandwidth, contrasting with vLLM's fine-grained 64-token paging.
To evaluate performance across different tiered storage systems, we configure the following four baselines:
(1) HBM: the standard vLLM serving with HBM only;
(2) DRAM (LMCache-DRAM-LW): extends capacity using host memory and applies \textit{layer-wise compute-I/O pipelining} to overlap retrieval overhead;
(3) LMCache-SSD: offloads KV data to NVMe SSDs using memcopy and standard asynchronous I/O;
and (4) LMCache-GDS: further optimizes SSD access using GDS to bypass the CPU bounce buffer.
Unless otherwise stated in end-to-end results, DRAM refers to LMCache-DRAM-LW; in ablations, we additionally report LMCache-DRAM without layerwise transfer.

\noindent\textbf{Models.}
We primarily evaluate performance using the Llama3-8B\cite{llama3-8B} model on a single GPU. 
To assess the scalability of our system in ultra-long sequence inference, we additionally employ GLM-4-9B-Chat-1M\cite{glm}. 
This model, which supports a 1M token context window, is distributed across two GPUs using Tensor Parallelism.

\noindent\textbf{Workloads.}
We use two established benchmarks: LEval\cite{leval} and LooGLE\cite{loogle}. 

LEval is a comprehensive long-context evaluation suite comprising 20 sub-tasks categorized into two main groups, covering a wide range of domains, including law, finance, technology, academic papers, and code.
The input lengths in LEval span a broad spectrum from 3k to 200k tokens. 
LooGLE, including 4 sub-tasks, is tailored for ultra-long context understanding, featuring significantly higher average document lengths, with many test samples exceeding 100k tokens. 
It focuses on complex tasks such as long dependency QA and single-turn summarization. 

\begin{table}[t]
    \centering
    \renewcommand{\arraystretch}{0.9}
    \caption{Cache hit rates across different storage tiers.}
    \label{tab:storage_hit_rate}
    \begin{tabular}{lcc}
    \toprule
    \multirow{2}{*}{\textbf{Storage Medium}} & \multicolumn{2}{c}{\textbf{Cache Hit Rate (\%)}} \\
    \cmidrule(lr){2-3}
     & \textbf{LEval} & \textbf{LooGLE} \\
    \midrule
    HBM    & 8  & 4  \\
    DRAM   & 53 & 24 \\
    SSD    & 84 & 86 \\
    \bottomrule
    \end{tabular}
\end{table}

Under our current system configuration, cache hit rates across storage tiers are shown in Table~\ref{tab:storage_hit_rate}.
HBM capacity is insufficient for long-context serving, yielding only 8\% and 4\% hit rates on LEval and LooGLE, respectively.
DRAM improves reuse to 53\% (LEval) and 24\% (LooGLE), while the larger context lengths in LooGLE still cause substantial misses.
In contrast, SSD sustains consistently high hit rates (84\% and 86\%), indicating that most reusable KV states can be captured by the large-capacity SSD tier.

To simulate a multi-session environment, we adopt a round-robin strategy to extract requests from the various sub-datasets of LEval and LooGLE.
In order to assess system robustness under varying load conditions, we simulate query arrivals via a Poisson distribution, as the datasets lack native timestamps. 
This setup aligns with the evaluation protocols adopted in prior works\cite{mooncake,vllm}.
These requests are continuously pushed into the vLLM serving engine, mimicking a real-world scenario where multiple users concurrently submit diverse queries with varying context lengths.

\noindent\textbf{Metrics.}
We evaluate \sys using two categories of metrics: end-to-end application performance and system-level micro-benchmarks.
We focus on two standard serving latencies:
(1) TTFT, which measures the responsiveness of the prefill phase;
and (2) ITL, which quantifies the decoding speed.
We report average latency under concurrent load.

To dissect the contributions of our system components, we measure:
(1) Cache Hit Rate, specifically analyzing its impact on reducing TTFT; 
(2) Storage Bandwidth, to evaluate the raw throughput of our storage engine; 
(3) GPU Bubble Time, to assess the efficacy of our asynchronous I/O scheduling in hiding latency;
and (4) Inference Cost, to evaluate the cost-effectiveness of our design.

\subsection{End-to-End Performance} \label{sec:e2e}

As illustrated in Figure~\ref{fig:e2e}, \sys demonstrates end-to-end performance and stability compared to all baselines.
With the newer software version (vLLM 0.17.0), \sys still delivers the best end-to-end latency across both workloads, confirming that our storage-compute co-design remains effective even when the serving engine becomes more compute-efficient.
%
%

\textbf{Time to First Token.} 
Across both software generations, HBM and SSD baselines remain weak for TTFT.
HBM is constrained by limited capacity and low hit rates, which triggers frequent KV recomputation.
SSD is constrained by longer I/O latency and CPU-side software overheads (e.g., memory allocation/release), which increase I/O jitter and queueing delay, and in turn enlarge GPU stall time.
On LEval with the old version, DRAM, GDS, and \sys all provide usable TTFT at high request rates (RPS, requests per second), while \sys remains the best and improves over GDS by 71.8\% at the highest load point.
With the new version, compute becomes faster and the relative cost of the GDS I/O path becomes more visible; at high load, DRAM now reduces TTFT by 29.6\% compared with GDS.
Even under this shift, \sys stays optimal, reducing TTFT by 69.1\% versus DRAM and 78.3\% versus GDS.
Under a 1s TTFT SLO, \sys increases the effective request rate by 50\% over DRAM and by 100\% over GDS.
On LooGLE, the longer requests make HBM and SSD consistently poor in both versions.
In the old version, GDS still provides clear benefits over DRAM and is relatively closer to \sys.
In the new version, GDS continues to outperform DRAM but its relative benefit decreases, and at 0.6 RPS its TTFT is still about 2.63$\times$ that of \sys.
At the same load point, \sys reduces TTFT by 93.2\% versus DRAM and 62.0\% versus GDS.
%

\textbf{Inter-Token Latency.}
In the old version on LEval, \sys already outperformed both DRAM and GDS at high load: at 1.5 RPS, ITL is reduced by 60.4\% versus DRAM and 24.9\% versus GDS.
In the new version, \sys remains the best decode path; at 1.5 RPS on LEval, ITL is still reduced by 22.0\% versus DRAM and 24.4\% versus GDS.
The gain comes from two effects: \sys provides higher effective cache hits during decode and reduces the compute-I/O gap, so the GPU spends more cycles on useful token generation instead of waiting for data.
On LooGLE, the gain in the new version narrows (18.3\% over GDS at 0.5 RPS and 10.2\% at 0.6 RPS), but \sys remains consistently better.
The gap narrows on LooGLE because much longer inputs increase per-token compute time, making decode relatively more compute-dominated.
Even with this narrowing, \sys maintains the lowest and smoothest ITL curve, suggesting potential headroom to sustain higher RPS under the same ITL target.

\subsection{Ablations}

In this subsection, we conduct ablation studies to isolate the contribution of key design components in \sys.
We evaluate five aspects: raw retrieve/store bandwidth, PRP vs SGL command path, TTFT under varying prefix reuse, distributed scalability, and the effectiveness of layerwise asynchronous pipelining.
These ablations directly evaluate key elements of our GPU-centric object-storage path, including command submission overheads, transfer bandwidth, and overlap efficiency.
To make the DRAM baselines explicit in this section, we additionally report both LMCache-DRAM and LMCache-DRAM-LW.
LMCache-DRAM denotes the DRAM path without layerwise (LW) copy/overlap, while LMCache-DRAM-LW denotes the DRAM path with layerwise memory copy and overlap.

\subsubsection{Bandwidth Performance of Retrieve and Store}

\begin{figure}
    \centering
     \includegraphics[width=1\linewidth]{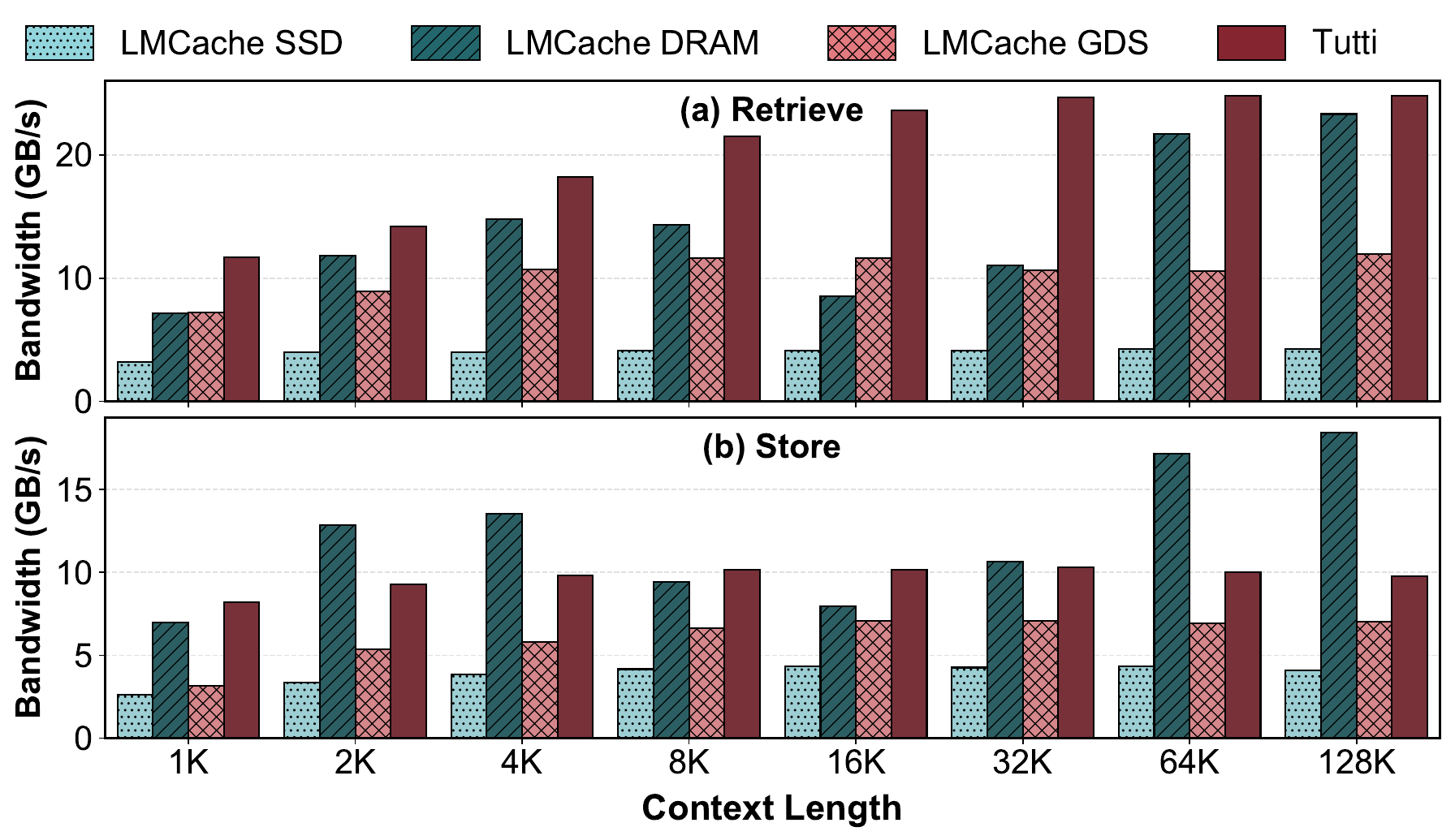}
    \caption{Raw bandwidth of retrieve and store interfaces across varying context lengths. }
    \label{fig:ablation}
\end{figure}

To isolate the performance characteristics of the storage subsystem, we bypass the model execution pipeline and directly benchmark the raw bandwidth of the \texttt{retrieve} and \texttt{store} interfaces.
All SSD-based backends are evaluated using a two-disk RAID-0 configuration. 
Evaluations cover a range of sequence lengths from 1K to 128K tokens across four representative storage backends.
As the prefix length increases, retrieval bandwidth emerges as the dominant performance factor, as illustrated in Figure~\ref{fig:ablation}(a).
LMCache-DRAM exhibits significant instability—for example, its throughput drops to 8.5 GB/s at 16K tokens due to memory fragmentation overhead. 
In contrast, \sys maintains a smooth, near-linear scaling trend, reaching up to 25.9 GB/s for longer contexts.
Compared to LMCache-GDS, whose performance saturates at around 11.9 GB/s even with two SSDs, \sys achieves up to a 2.08$\times$ higher retrieval bandwidth. 
Figure~\ref{fig:ablation}(b) reports the store bandwidth.
While LMCache-DRAM naturally reaches the highest raw bandwidth (up to 18.4 GB/s) thanks to in-memory writes, it lacks persistence and is limited by DRAM capacity.
Among persistent storage backends, \sys consistently outperforms both LMCache-SSD and LMCache-GDS: it sustains roughly 10 GB/s write bandwidth across all tested lengths (e.g., 9.8 GB/s at 128K tokens), whereas LMCache-GDS remains around 7 GB/s despite using the same dual-SSD configuration.
Notably, \sys performance is constrained by the storage device itself, as each SSD provides no more than 10 GB/s peak sequential store bandwidth.     
Prior work~\cite{ren2025characterizing} indicates that store bandwidth is less critical than retrieval bandwidth for end-to-end inference performance, and 10 GB/s is sufficient to sustain high performance in most scenarios.

\begin{figure}
    \centering
     \includegraphics[width=0.8\linewidth]{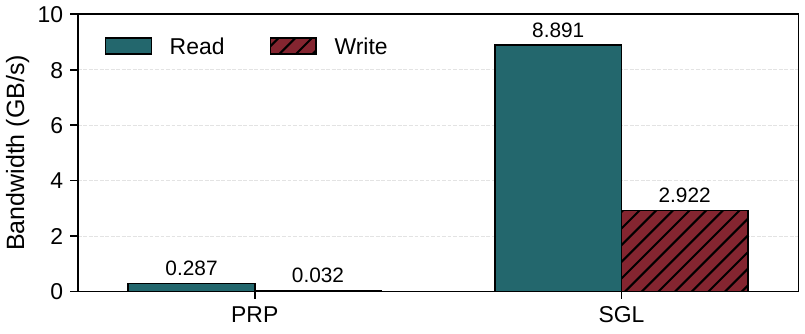}
    \caption{PRP vs SGL bandwidth under a single-thread read/write microbenchmark. Compared with PRP, SGL delivers substantially higher read and write bandwidth.}
    \label{fig:prpsgl_bandwidth}
\end{figure}

\subsubsection{PRP vs SGL Bandwidth.}

To validate the impact of applying SGL in our design, we run a single-GPU-thread microbenchmark that reads and writes 500 MB of data per operation.
As shown in Figure~\ref{fig:prpsgl_bandwidth}, under PRP the read/write bandwidth is 0.287 GB/s and 0.032 GB/s, while switching to SGL improves it to 8.891 GB/s and 2.922 GB/s, corresponding to 31.0$\times$ and 91.3$\times$ gains.
The key reason is that SGL commands reduce PCIe communication overhead between host and NVMe devices compared with PRP, which lowers command/descriptor handling overhead and stabilizes queue progress.

\begin{figure}[t]
    \centering
     \includegraphics[width=1\linewidth]{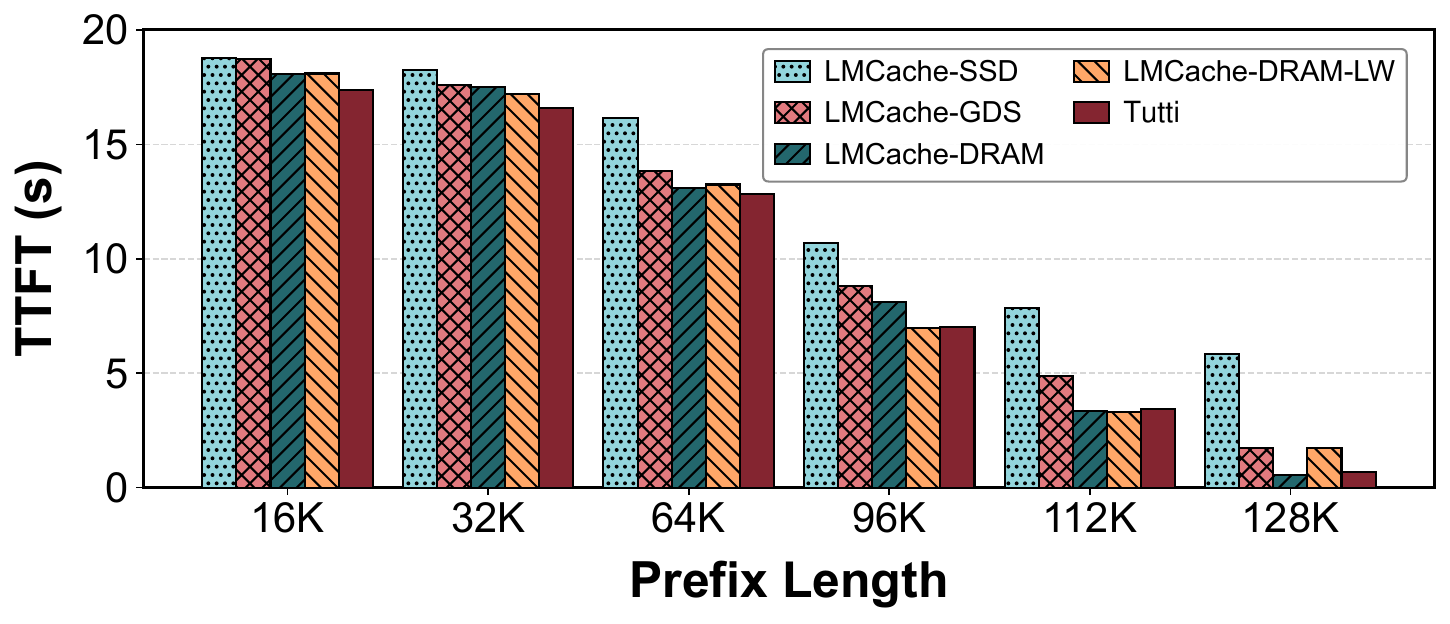}
     \caption{TTFT performance comparison across varying prefix lengths on Llama3-8B-Instruct (Single-GPU). 
     \sys demonstrates superior I/O efficiency, achieving up to 61.4\% lower TTFT than LMCache-GDS.}
    \label{fig:ablation_ttft}
\end{figure}

\begin{figure}[t]
    \centering
     \includegraphics[width=1\linewidth]{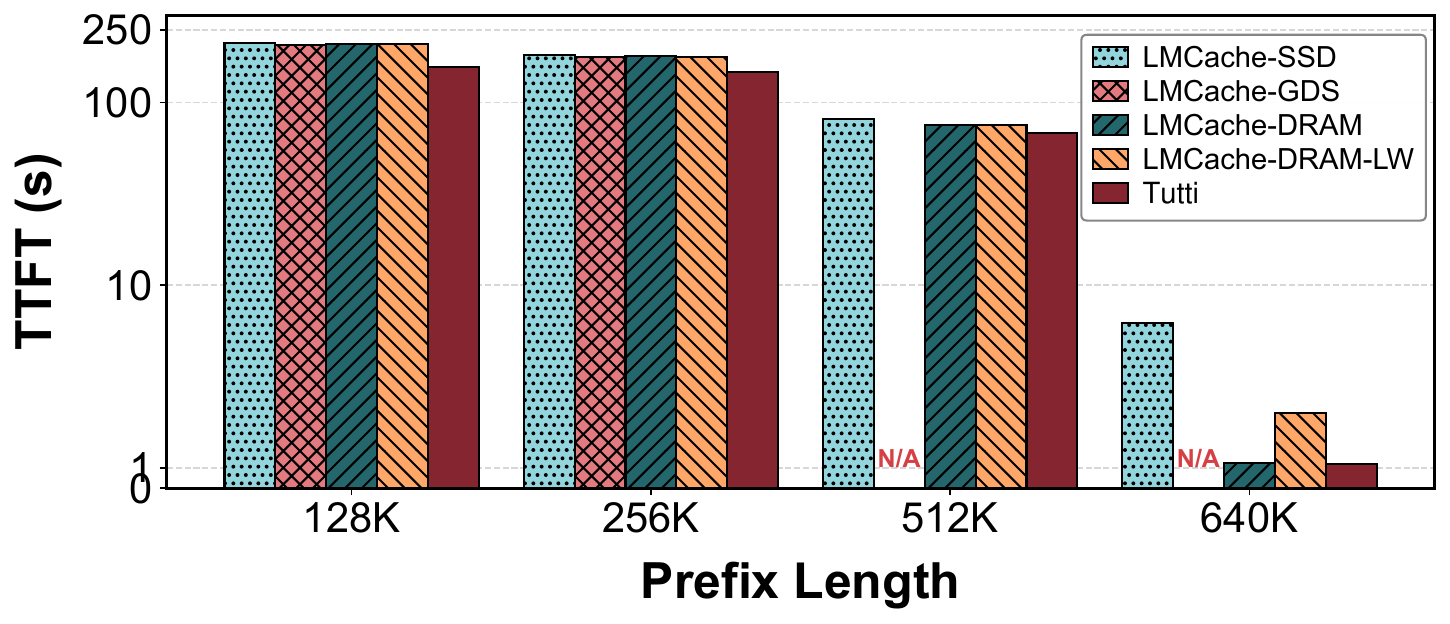}
     \caption{Distributed Scalability for GLM-4-9B-1M (2-GPU, 4-Disk). 
     \sys overcomes LMCache-GDS's OOM failure at 512K/640K by avoiding staging buffer overhead, demonstrating architectural robustness and achieving the best TTFT 1.2s at 640K.}
    \label{fig:glm_ttft}
\end{figure}

\subsubsection{TTFT Performance across Context Lengths.}
We evaluate TTFT under varying prefix reuse by fixing the total input length to 128k tokens and increasing the cached prefix from 16k to 128k. 
LMCache-SSD suffers severe degradation under high reuse due to limited bandwidth; at a 112k prefix, its TTFT rises to 7.84s. 
In contrast, our system sustains stable performance by overlapping retrieval with the remaining computation, achieving 3.43s at the same prefix—2.28$\times$ faster than SSD.
Compared to LMCache-GDS, our method consistently maintains an advantage across all prefix lengths, with improvements ranging from 5.8\% at 32k up to 61.4\% at 128k. 
Notably, for moderate reuse (16k–96k), our system even matches or exceeds DRAM performance, achieving up to 13.4\% improvement—indicating that effective I/O–compute overlap can outweigh DRAM’s raw latency. 
Only in extremely high reuse conditions (>96k), where the workload becomes almost purely retrieval-bound, does DRAM regain its expected lead, with our system trailing by at most 20.6\%.

\subsubsection{Multi-GPU Scalability} \label{eva-gds-oom}
In order to evaluate the scalability of \sys in distributed settings, we test the TTFT performance using the GLM-4-9B-Chat-1M model across two GPUs (each residing under a PCIe Root complex) and four disks (two attached to each GPU's root complex). 
GPUs are connected via NVLink.

The experimental data highlights the superior performance of \sys: for a 128K Prefix Length, \sys achieved a TTFT of only 155.743 s, representing an approximate 25\% latency reduction compared to LMCache-GDS (207.12 s). 
LMCache-GDS in longer contexts exposes a critical limitation stemming from its reliance on GDS technology. 
GDS leverages the cufile to achieve direct data transfer from storage to GPU memory. 
Crucially, to enable this mechanism outside the inference work, cufile must allocate a certain block of GPU memory to serve as a staging buffer. 
In long inference, this memory allocation for I/O acceleration quickly exceeded the available GPU memory capacity, triggering a fatal Out-of-Memory (OOM) error. 
Consequently, LMCache-GDS failed to complete the tests at both 512K and 640K (marked as N/A). 
In contrast, \sys deeply integrates with the inference engine and provides register interfaces to directly manage GPU memory without the need for an intermediate staging buffer.
This allowed \sys to successfully complete the most challenging tests, ultimately achieving the overall best TTFT of 1.2 seconds at the extreme 640K Prefix Length.
Our perspective is that high-performance I/O must be deeply integrated and co-optimized with computation, instead of being treated as a simple third-party plugin.

\subsubsection{Comparison of Layerwise Async Pipelining}
\begin{figure}
    \centering
     \includegraphics[width=1\linewidth]{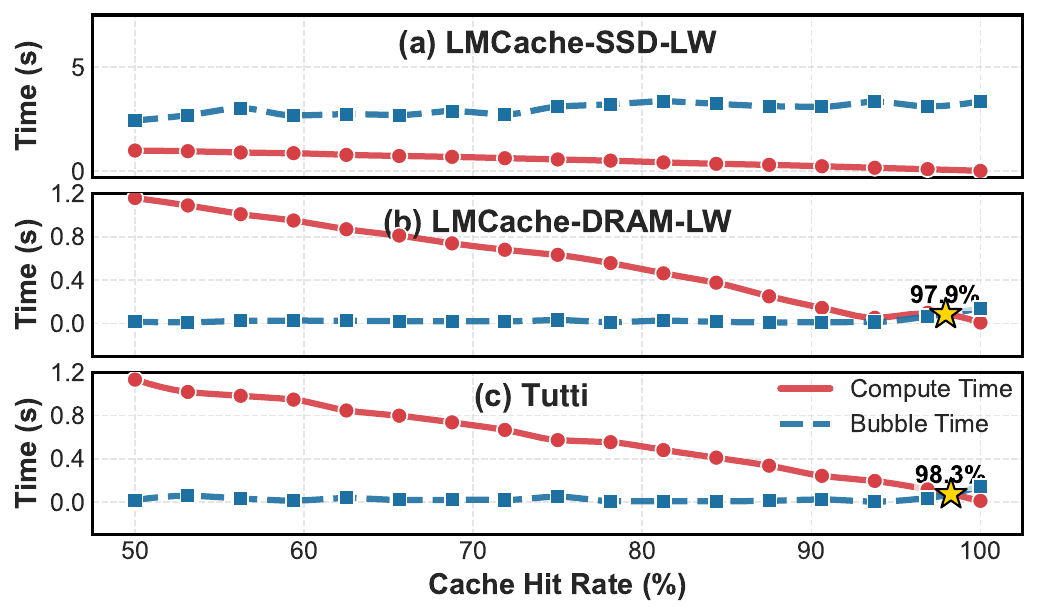}
     \caption{
     Decomposition of latency by cache hit rate, highlighting the critical Crossover Point ($\star$) where bubble time begins to exceed compute time. 
     Our layerwise asynchronous mechanism successfully pushes this critical point to an extremely high cache hit rate of 98.3\%, maintaining a near-optimal compute-bound across the tested range.
     }
     
    \label{fig:ablation_bubble}
\end{figure}
\label{sec:ablation_async}

To verify the effectiveness of the Slack-Aware I/O Scheduler, we break down the total inference latency into computation time and bubble time. 
This evaluation compares the performance profiles across three distinct storage backends, all of which utilize a layerwise pipelining strategy.
We specifically exclude LMCache-GDS from this latency decomposition study, as its current implementation does not support this strategy.
By fixing the prompt length at 32K and varying the hit rate, we manipulate the compute-to-load ratio.
The core principle is to achieve a deep overlap between data transmission and layerwise computation.
Ideally, as long as the computation time for a layer exceeds its data transfer time ($T_{compute} > T_{transfer}$), the transmission latency can be completely masked, resulting in near-zero bubble time.

As illustrated in Figure~\ref{fig:ablation_bubble}(a), the LMCache-SSD bubble is excessive; it cannot be effectively hidden by the shorter computation phases.
The inefficient pipeline exposes raw transfer latency, resulting in substantial bubble time (blue dashed line) and degraded end-to-end performance.
In contrast, Figure~\ref{fig:ablation_bubble}(c) demonstrates that our system successfully masks transmission overhead. 
Across the majority of the testing range, bubble time is negligible (averaging ~25ms), and drops to a mere 6ms at a 93.75\% hit rate.
\sys maintains a compute-bound profile (dominated by the red solid line), achieving a near-optimal execution curve that is similar to the DRAM-based strong baseline shown in Figure~\ref{fig:ablation_bubble}(b). 
We further identify a critical "crossover point" (marked by a star), which indicates the transition from a compute-bound to an I/O-bound state.  
Critically, for our system, this crossover point is pushed to an extremely high cache hit rate of 98.3\%, a significant improvement compared to the much lower thresholds observed in the LMCache-SSD baseline. 
This result definitively proves that our layerwise mechanism successfully extends the "effective zero-bubble zone" to its physical limits, only introducing minor bubbles when the computation becomes exceedingly sparse.

\begin{figure}
    \centering
    \includegraphics[width=1\linewidth]{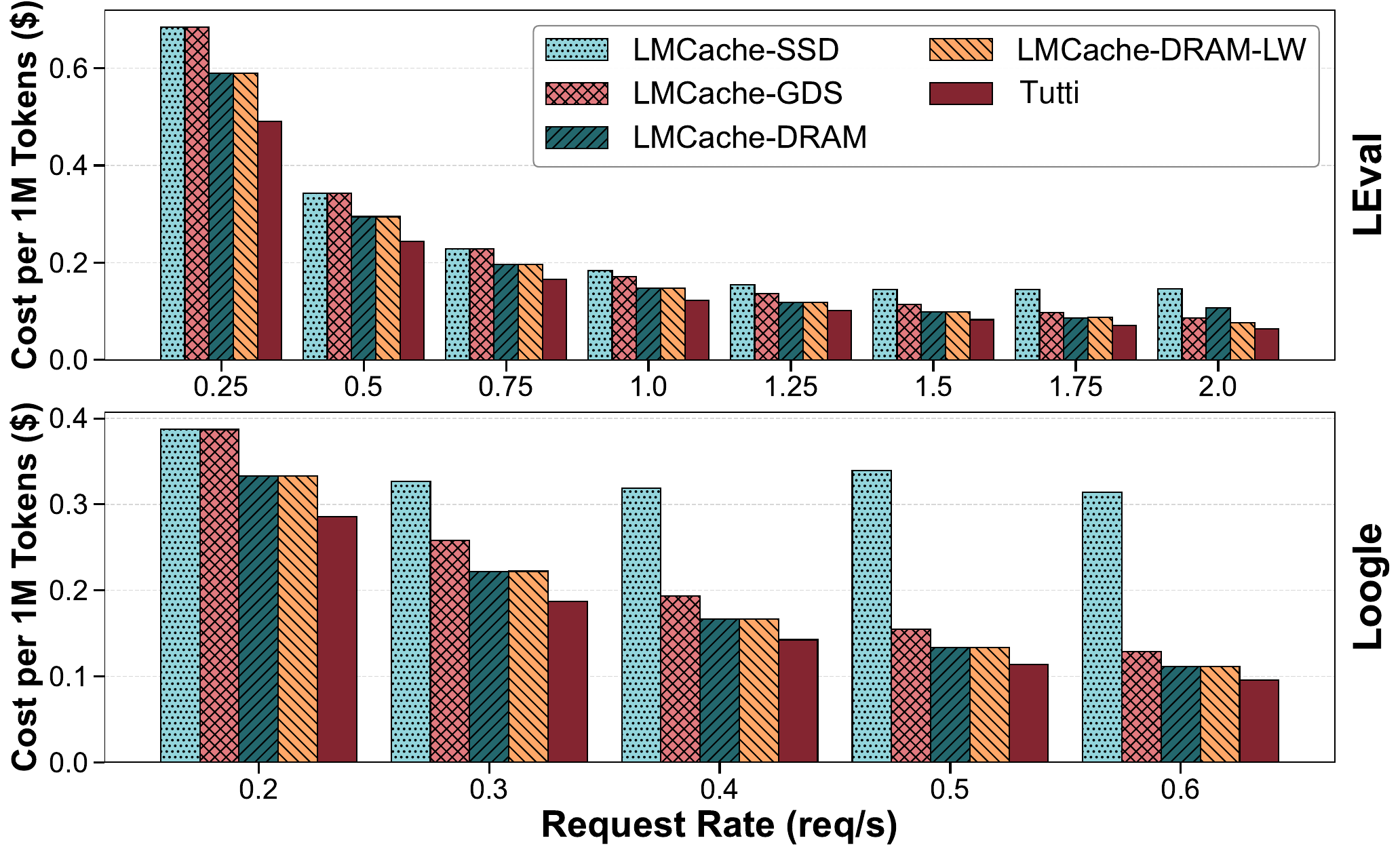}
     \caption{Inference cost per 1 million tokens across LEval and LooGLE workloads. 
     \sys achieves the lowest cost by leveraging SSDs.
}
    \label{fig:cost_analysis}
\end{figure}

\subsection{Inference Cost} 

To quantify the economic benefits of \sys, we calculate the serving cost normalized by the token generation throughput.
The total cost aggregates the expenses of GPU and the tiered storage hierarchy (DRAM and SSD). 
The formula is defined as:
\begin{equation}
    \text{Cost}_{1M} = \frac{
        \overbrace{P_{GPU} \cdot N_{GPU}}^{\text{Compute Cost}} + 
        \overbrace{P_{mem} \cdot S_{mem} + P_{ssd} \cdot S_{ssd}}^{\text{Storage Cost}}
    }{
        \text{Throughput (tokens/hour)}
    } \times 10^6
\end{equation}
\noindent where $P_{GPU}$ is the hourly GPU price, $N_{GPU}$ is the GPU count, and $P_{x}/S_{x}$ represent the unit price and capacity for DRAM/SSD, respectively.

We adopt typical cloud pricing: \$5/hour per NVIDIA H100 GPU, \$0.0088/GB/hour for DRAM, and \$0.000082/GB/hour for NVMe SSD~\cite{ec2_p4d_pricing, ec2_pricing}.
Figure~\ref{fig:cost_analysis} illustrates the cost per 1 million tokens for the LEval and LooGLE workloads.
\sys consistently demonstrates the most favorable cost-efficiency profile across all request rates.
With increasing context lengths, DRAM-based systems are forced to provision larger memory capacities for the KV cache, resulting in significantly higher operational costs.
In contrast, \sys offloads the majority of KV data to SSDs (which are approximately $100\times$ cheaper per GB than DRAM).
While LMCache-SSD leverages the same cost-effective storage medium, its inherent performance overheads bottleneck throughput. 
This inefficiency leads to GPU underutilization, effectively inflating the unit cost. 
In contrast, our system fully saturates GPU compute resources, maximizing throughput and optimizing the yield of tokens per GPU-hour.
Specifically, on the LooGLE workload at 0.5 QPS, \sys reduces the serving cost by 66.2\% compared to LMCache-SSD and outperforms LMCache-GDS by roughly 27\%.

\section{Conclusion and Future Work}
In this paper, we presented \sys, a GPU-centric, SSD-backed KV cache store for long-context LLM serving.
\sys removes CPU intervention from critical data and I/O control paths between GPU HBM and NVMe SSDs.
By combining a GPU-centric object-storage design with a layerwise GPU compute-I/O pipeline, \sys enables SSD-backed KV caching to achieve DRAM-like efficiency while effectively suppressing GPU stall time.
Our evaluation shows that, compared with the SOTA GDS-enabled SSD-backed solution, \sys reduces TTFT by 78.3\% under strict SLO constraints and improves the achievable request rate by 2$\times$.
\sys also lowers the LLM serving cost by about 27\%.

\begin{acks}
We would like to thank Menglei Chen from Huazhong University of Science and Technology for his valuable guidance on GPU hashing during the early stages of this work. We are also grateful to Zheng Zhang from Wuhan University for his guidance and assistance with performance profiling of the GPU I/O kernel.
\end{acks}
\bibliographystyle{ACM-Reference-Format}
\bibliography{references}

@misc{hicache,
  title        = {SGLang HiCache: Fast Hierarchical KV Caching with Your Favorite Storage Backends},
  author       = {{Xie, Zhiqiang}},
  year         = {2025},
  howpublished = {\url{https://lmsys.org/blog/2025-09-10-sglang-hicache/}},
}

@misc{ec2_pricing,
  title        = {Amazon ec2 pricing.},
  author       = {{AWS}},
  year         = {2025},
  howpublished = {\url{https://aws.amazon.com/ec2/pricing/}},
}

@misc{ec2_p4d_pricing,
  title        = {Amazon ec2 p4d pricing.},
  author       = {{AWS}},
  year         = {2025},
  howpublished = {\url{https://aws.amazon.com/ec2/instance-types/p4/}},
}

@misc{llama3-8B,
  title        = {Meta-Llama-3-8B-Instruct},
  author       = {{Meta AI}},
  year         = {2024},
  howpublished = {\url{https://huggingface.co/meta-llama/Meta-Llama-3-8B-Instruct}},
  note         = {Accessed: 2025-12-10},
  publisher    = {Hugging Face}
}

@article{glm,
  title={Chatglm: A family of large language models from glm-130b to glm-4 all tools},
  author={GLM, Team and Zeng, Aohan and Xu, Bin and Wang, Bowen and Zhang, Chenhui and Yin, Da and Zhang, Dan and Rojas, Diego and Feng, Guanyu and Zhao, Hanlin and others},
  journal={arXiv preprint arXiv:2406.12793},
  year={2024}
}

@misc{D7-PS1010,
  author = {solidigm},
  title = {Solidigm D7-PS1010},
  year = {2024},
  howpublished = {\url{https://www.solidigm.com/products/data-center/d7/ps1010.html}}
}

@misc{loogle,
      title={LooGLE: Can Long-Context Language Models Understand Long Contexts?}, 
      author={Jiaqi Li and Mengmeng Wang and Zilong Zheng and Muhan Zhang},
      year={2024},
      eprint={2311.04939},
      archivePrefix={arXiv},
      primaryClass={cs.CL},
      url={https://arxiv.org/abs/2311.04939}, 
}

@misc{leval,
      title={L-Eval: Instituting Standardized Evaluation for Long Context Language Models}, 
      author={Chenxin An and Shansan Gong and Ming Zhong and Xingjian Zhao and Mukai Li and Jun Zhang and Lingpeng Kong and Xipeng Qiu},
      year={2023},
      eprint={2307.11088},
      archivePrefix={arXiv},
      primaryClass={cs.CL},
      url={https://arxiv.org/abs/2307.11088}, 
}

@article{gemini_1.5,
  title={Gemini 1.5: Unlocking multimodal understanding across millions of tokens of context},
  author={Team, Gemini and Georgiev, Petko and Lei, Ving Ian and Burnell, Ryan and Bai, Libin and Gulati, Anmol and Tanzer, Garrett and Vincent, Damien and Pan, Zhufeng and Wang, Shibo and others},
  journal={arXiv preprint arXiv:2403.05530},
  year={2024}
}

@article{mooncake,
  title   = {Mooncake: Kimi's KVCache-centric Architecture for LLM Serving},
  author  = {Qin, Ruoyu and Li, Zheming and He, Weiran and Zhang, Mingxing and Wu, Yongwei and Zheng, Weimin and Xu, Xinran},
  journal = {arXiv preprint arXiv:2407.00079},
  year    = {2024}
}

@inproceedings{cacheblend,
author = {Yao, Jiayi and Li, Hanchen and Liu, Yuhan and Ray, Siddhant and Cheng, Yihua and Zhang, Qizheng and Du, Kuntai and Lu, Shan and Jiang, Junchen},
title = {CacheBlend: Fast Large Language Model Serving for RAG with Cached Knowledge Fusion},
year = {2025},
isbn = {9798400711961},
publisher = {Association for Computing Machinery},
address = {New York, NY, USA},
url = {https://doi.org/10.1145/3689031.3696098},
doi = {10.1145/3689031.3696098},
booktitle = {Proceedings of the Twentieth European Conference on Computer Systems},
pages = {94–109},
numpages = {16},
location = {Rotterdam, Netherlands},
series = {EuroSys '25}
}

@misc{per-token-cost-deepseek,
  author = {Deepseek},
  title = {Models and Pricing},
  year = {2025},
  howpublished = {\url{https://api-docs.deepseek.com/quick_start/pricing/}}
}

@misc{per-token-cost-openai,
  author = {OpenAI},
  title = {API Pricing},
  year = {2025},
  howpublished = {\url{https://openai.com/api/pricing/}}
}

@inproceedings{cacheattention,
  title={Cost-Efficient large language model serving for multi-turn conversations with CachedAttention},
  author={Gao, Bin and He, Zhuomin and Sharma, Puru and Kang, Qingxuan and Jevdjic, Djordje and Deng, Junbo and Yang, Xingkun and Yu, Zhou and Zuo, Pengfei},
  booktitle={2024 USENIX Annual Technical Conference (USENIX ATC 24)},
  pages={111--126},
  year={2024}
}

@inproceedings{IMPRESS,
  title={IMPRESS: An Importance-InformedMulti-Tier Prefix KV Storage System for Large Language Model Inference},
  author={Chen, Weijian and He, Shuibing and Qu, Haoyang and Zhang, Ruidong and Yang, Siling and Chen, Ping and Zheng, Yi and Huai, Baoxing and Chen, Gang},
  booktitle={23rd USENIX Conference on File and Storage Technologies (FAST 25)},
  pages={187--201},
  year={2025}
}

@inproceedings{geminifs,
  title={GeminiFS: A Companion File System for GPUs},
  author={Qiu, Shi and Liu, Weinan and Hu, Yifan and Yan, Jianqin and Shen, Zhirong and Yao, Xin and Chen, Renhai and Zhang, Gong and Zhang, Yiming},
  booktitle={23rd USENIX Conference on File and Storage Technologies (FAST 25)},
  pages={221--236},
  year={2025}
}

@inproceedings{pagedattention,
  title={Efficient memory management for large language model serving with pagedattention},
  author={Kwon, Woosuk and Li, Zhuohan and Zhuang, Siyuan and Sheng, Ying and Zheng, Lianmin and Yu, Cody Hao and Gonzalez, Joseph and Zhang, Hao and Stoica, Ion},
  booktitle={Proceedings of the 29th symposium on operating systems principles},
  pages={611--626},
  year={2023}
}

@inproceedings{code_understanding,
  title={Using an llm to help with code understanding},
  author={Nam, Daye and Macvean, Andrew and Hellendoorn, Vincent and Vasilescu, Bogdan and Myers, Brad},
  booktitle={Proceedings of the IEEE/ACM 46th International Conference on Software Engineering},
  pages={1--13},
  year={2024}
}

@article{multi_turn_dialogue_systems,
  title={A survey on recent advances in llm-based multi-turn dialogue systems},
  author={Yi, Zihao and Ouyang, Jiarui and Liu, Yuwen and Liao, Tianhao and Xu, Zhe and Shen, Ying},
  journal={arXiv preprint arXiv:2402.18013},
  year={2024}
}

@article{vaswani2017attention,
  title={Attention is all you need},
  author={Vaswani, Ashish and Shazeer, Noam and Parmar, Niki and Uszkoreit, Jakob and Jones, Llion and Gomez, Aidan N and Kaiser, {\L}ukasz and Polosukhin, Illia},
  journal={Advances in neural information processing systems},
  volume={30},
  year={2017}
}

@inproceedings{vllm,
  title={Efficient Memory Management for Large Language Model Serving with PagedAttention},
  author={Woosuk Kwon and Zhuohan Li and Siyuan Zhuang and Ying Sheng and Lianmin Zheng and Cody Hao Yu and Joseph E. Gonzalez and Hao Zhang and Ion Stoica},
  booktitle={Proceedings of the ACM SIGOPS 29th Symposium on Operating Systems Principles},
  year={2023}
}

@misc{SGLang,
  author = {SGLang},
  title = {SGLang},
  year = {2024},
  howpublished = {\url{https://github.com/sgl-project/sglang?tab=readme-ov-file}}
}

@inproceedings{hcache,
  title={Fast state restoration in llm serving with hcache},
  author={Gao, Shiwei and Chen, Youmin and Shu, Jiwu},
  booktitle={Proceedings of the Twentieth European Conference on Computer Systems},
  pages={128--143},
  year={2025}
}

@inproceedings{lmcache,
  title={Cachegen: Kv cache compression and streaming for fast large language model serving},
  author={Liu, Yuhan and Li, Hanchen and Cheng, Yihua and Ray, Siddhant and Huang, Yuyang and Zhang, Qizheng and Du, Kuntai and Yao, Jiayi and Lu, Shan and Ananthanarayanan, Ganesh and others},
  booktitle={Proceedings of the ACM SIGCOMM 2024 Conference},
  pages={38--56},
  year={2024}
}

@inproceedings{ren2025characterizing,
  title={An I/O Characterizing Study of Offloading LLM Models and KV Caches to NVMe SSD},
  author={Ren, Zebin and Doekemeijer, Krijn and De Matteis, Tiziano and Pinto, Christian and Stoica, Radu and Trivedi, Animesh},
  booktitle={Proceedings of the 5th Workshop on Challenges and Opportunities of Efficient and Performant Storage Systems},
  pages={23--33},
  year={2025}
}

@inproceedings{lee2025disk,
  title={Disk-Based Shared KV Cache Management for Fast Inference in Multi-Instance LLM RAG Systems},
  author={Lee, Hyungwoo and Kim, Kihyun and Kim, Jinwoo and So, Jungmin and Cha, Myung-Hoon and Kim, Hong-Yeon and Kim, James J and Kim, Youngjae},
  booktitle={2025 IEEE 18th International Conference on Cloud Computing (CLOUD)},
  pages={199--209},
  year={2025},
  organization={IEEE}
}

@inproceedings{bam,
  title     = {GPU-initiated on-demand high-throughput storage access in the BaM system architecture},
  author    = {Qureshi, Zaid and Mailthody, Vikram Sharma and Gelado, Isaac and Min, Seungwon and Masood, Amna and Park, Jeongmin and Xiong, Jinjun and Newburn, Chris J and Vainbrand, Dmitri and Chung, I-Hsin and others},
  booktitle = {Proceedings of the 28th ACM International Conference on Architectural Support for Programming Languages and Operating Systems, Volume 2},
  pages     = {325--339},
  year      = {2023}
}

@inproceedings{gmt,
  title     = {GMT: GPU Orchestrated Memory Tiering for the Big Data Era},
  author    = {Chang, Chia-Hao and Han, Jihoon and Sivasubramaniam, Anand and Sharma Mailthody, Vikram and Qureshi, Zaid and Hwu, Wen-Mei},
  booktitle = {Proceedings of the 29th ACM International Conference on Architectural Support for Programming Languages and Operating Systems, Volume 3},
  pages     = {464--478},
  year      = {2024}
}

@inproceedings{pan2025instattention,
  title={InstAttention: In-Storage Attention Offloading for Cost-Effective Long-Context LLM Inference},
  author={Pan, Xiurui and Li, Endian and Li, Qiao and Liang, Shengwen and Shan, Yizhou and Zhou, Ke and Luo, Yingwei and Wang, Xiaolin and Zhang, Jie},
  booktitle={2025 IEEE International Symposium on High Performance Computer Architecture (HPCA)},
  pages={1510--1525},
  year={2025},
  organization={IEEE}
}

@misc{GDS,
  author       = {Nvidia},
  year         = {2024},
  title        = {NVIDIA GPUDirect Storage.},
  howpublished = {\url{https://docs.nvidia.com/gpudirect-storage/index.html}}
}

@misc{llama4,
  author = {Meta},
  title = {The Llama 4 herd: The beginning of a new era of natively multimodal AI innovation},
  year = {2025},
  howpublished = {\url{https://ai.meta.com/blog/llama-4-multimodal-intelligence/}}
}

@article{jiang2024kvpr,
  title={KVPR: Efficient LLM Inference with I/O-Aware KV Cache Partial Recomputation},
  author={Jiang, Chaoyi and Gao, Lei and Zarch, Hossein Entezari and Annavaram, Murali},
  journal={arXiv preprint arXiv:2411.17089},
  year={2024}
}

@article{gao2024attentionstore,
  title={Attentionstore: Cost-effective attention reuse across multi-turn conversations in large language model serving},
  author={Gao, Bin and He, Zhuomin and Sharma, Puru and Kang, Qingxuan and Jevdjic, Djordje and Deng, Junbo and Yang, Xingkun and Yu, Zhou and Zuo, Pengfei},
  journal={arXiv preprint arXiv:2403.19708},
  volume={52},
  pages={20--38},
  year={2024}
}

@article{ye2024chunkattention,
  title={Chunkattention: Efficient self-attention with prefix-aware kv cache and two-phase partition},
  author={Ye, Lu and Tao, Ze and Huang, Yong and Li, Yang},
  journal={arXiv preprint arXiv:2402.15220},
  year={2024}
}

@article{zheng2024sglang,
  title={Sglang: Efficient execution of structured language model programs},
  author={Zheng, Lianmin and Yin, Liangsheng and Xie, Zhiqiang and Sun, Chuyue Livia and Huang, Jeff and Yu, Cody Hao and Cao, Shiyi and Kozyrakis, Christos and Stoica, Ion and Gonzalez, Joseph E and others},
  journal={Advances in neural information processing systems},
  volume={37},
  pages={62557--62583},
  year={2024}
}

@misc{TensorRT-LLM,
  author = {NIVDIA},
  title = { TensorRT LLM’s Documentation},
  year = {2025},
  howpublished = {\url{https://nvidia.github.io/TensorRT-LLM/}}
}

@misc{DeepNVMe,
  author = {DeepSpeedAI},
  title = {DeepNVMe: Affordable I/O scaling for Deep Learning Applications},
  year = {2025},
  howpublished = {\url{https://github.com/deepspeedai/DeepSpeed/blob/master/blogs/deepnvme/06-2025/README.md/}}
}

@inproceedings{gofs,
  title={Managing Scalable Direct Storage Accesses for GPUs with GoFS},
  author={Li, Shaobo and Zhou, Yirui Eric and Xue, Yuqi and Xu, Yuan and Huang, Jian},
  booktitle={Proceedings of the ACM SIGOPS 31st Symposium on Operating Systems Principles},
  pages={979--995},
  year={2025}
}

@article{xie2025strata,
  title={Strata: Hierarchical Context Caching for Long Context Language Model Serving},
  author={Xie, Zhiqiang and Xu, Ziyi and Zhao, Mark and An, Yuwei and Mailthody, Vikram Sharma and Mahlke, Scott and Garland, Michael and Kozyrakis, Christos},
  journal={arXiv preprint arXiv:2508.18572},
  year={2025}
}

@article{shen2024fastswitch,
  title={Fastswitch: Optimizing context switching efficiency in fairness-aware large language model serving},
  author={Shen, Ao and Li, Zhiyao and Gao, Mingyu},
  journal={arXiv preprint arXiv:2411.18424},
  year={2024}
}

@article{qwen3,
    title={Qwen3 Technical Report}, 
    author={An Yang and Anfeng Li and Baosong Yang and Beichen Zhang and Binyuan Hui and Bo Zheng and Bowen Yu and Chang Gao and Chengen Huang and Chenxu Lv and Chujie Zheng and Dayiheng Liu and Fan Zhou and Fei Huang and Feng Hu and Hao Ge and Haoran Wei and Huan Lin and Jialong Tang and Jian Yang and Jianhong Tu and Jianwei Zhang and Jianxin Yang and Jiaxi Yang and Jing Zhou and Jingren Zhou and Junyang Lin and Kai Dang and Keqin Bao and Kexin Yang and Le Yu and Lianghao Deng and Mei Li and Mingfeng Xue and Mingze Li and Pei Zhang and Peng Wang and Qin Zhu and Rui Men and Ruize Gao and Shixuan Liu and Shuang Luo and Tianhao Li and Tianyi Tang and Wenbiao Yin and Xingzhang Ren and Xinyu Wang and Xinyu Zhang and Xuancheng Ren and Yang Fan and Yang Su and Yichang Zhang and Yinger Zhang and Yu Wan and Yuqiong Liu and Zekun Wang and Zeyu Cui and Zhenru Zhang and Zhipeng Zhou and Zihan Qiu},
    journal = {arXiv preprint arXiv:2505.09388},
    year={2025}
}

@article{storage-next,
  title={Advancing Memory and Storage Architectures for Next-Gen AI Workloads},
  author={Vikram Sharma Mailthody},
  journal={Flash Memory Summit},
  pages={1--27},
  year={2025}
}

@inproceedings{sheng2023flexgen,
  title={Flexgen: High-throughput generative inference of large language models with a single gpu},
  author={Sheng, Ying and Zheng, Lianmin and Yuan, Binhang and Li, Zhuohan and Ryabinin, Max and Chen, Beidi and Liang, Percy and R{\'e}, Christopher and Stoica, Ion and Zhang, Ce},
  booktitle={International Conference on Machine Learning},
  pages={31094--31116},
  year={2023},
  organization={PMLR}
}

@inproceedings{didona2022understandingio,
  title={Understanding modern storage APIs: a systematic study of libaio, SPDK, and io\_uring},
  author={Didona, Diego and Pfefferle, Jonas and Ioannou, Nikolas and Metzler, Bernard and Trivedi, Animesh},
  booktitle={Proceedings of the 15th ACM International Conference on Systems and Storage},
  pages={120--127},
  year={2022}
}

@misc{KIOXIA,
  author = {KIOXIA},
  title = {KIOXIA CM7-V Series Enterprise NVMe™ Mixed Use SSD},
  year = {2025},
  howpublished = {\url{https://apac.kioxia.com/en-apac/business/ssd/enterprise-ssd/cm7-v.html}}
}

@misc{Solidigm,
  author = {Solidigm},
  title = {Solidigm D7-PS1010},
  year = {2025},
  howpublished = {\url{https://www.solidigm.com/products/data-center/d7/ps1010.html\#configurator}}
}

@misc{nvme2,
  author       = {NVM Express Workgroup},
  year         = {2022},
  title        = {NVM Express Base Specification Revision 2.0c.},
  howpublished = {\url{https://nvmexpress.org/wp-content/uploads/NVM-Express-Base-Specification-2.0c-2022.10.04-Ratified.pdf}}
}

@article{lin2025bullet,
  title={Bullet: Boosting GPU Utilization for LLM Serving via Dynamic Spatial-Temporal Orchestration},
  author={Lin, Zejia and Xu, Hongxin and Chen, Guanyi and Zhang, Xianwei and Lu, Yutong},
  journal={arXiv preprint arXiv:2504.19516},
  year={2025}
}

@inproceedings{flashgen,
  title={Accelerating LLM Serving for Multi-turn Dialogues with Efficient Resource Management},
  author={Jeong, Jinwoo and Ahn, Jeongseob},
  booktitle={Proceedings of the 30th ACM International Conference on Architectural Support for Programming Languages and Operating Systems, Volume 2},
  pages={1--15},
  year={2025}
}

@misc{ali-kv,
  author       = {Aliyun},
  year         = {2025},
  title        = {PolarKVCache.},
  howpublished = {\url{https://help.aliyun.com/zh/polardb/polardb-for-mysql/user-guide/polarkvcache-inference-acceleration}}
}

@inproceedings{yan2025phoenix,
  title={Phoenix: A Refactored I/O Stack for GPU Direct Storage without Phony Buffers},
  author={Yan, Jianqin and Qiu, Shi and Lv, Yina and Hu, Yifan and Chen, Hao and Shen, Zhirong and Yao, Xin and Chen, Renhai and Shu, Jiwu and Zhang, Gong and others},
  booktitle={Proceedings of the International Conference for High Performance Computing, Networking, Storage and Analysis},
  pages={1267--1283},
  year={2025}
}

@misc{green-contex,
  author       = {NVIDIA},
  year         = {2025},
  title        = {Cuda-programming-guide Green Contexts.},
  howpublished = {\url{https://docs.nvidia.com/cuda/cuda-programming-guide/04-special-topics/green-contexts.html/}}
}

@misc{5min-tire,
  author       = {Doug O'Laughlin},
  year         = {2026},
  title        = {Another Conversation with Val Bercovici Memory Markets.},
  howpublished = {\url{https://www.fabricatedknowledge.com/p/another-conversation-with-val-bercovici}}
}

@misc{weka-kv,
  author       = {Devansh},
  year         = {2026},
  title        = {How Weka is Solving AI’s Trillion Dollar Memory Problem.},
  howpublished = {\url{https://www.artificialintelligencemadesimple.com/p/how-one-startup-is-breaking-nvidias?utm_source=publication-search}}
}

@article{smartio,
  title={Smartio: Zero-overhead device sharing through pcie networking},
  author={Markussen, Jonas and Kristiansen, Lars Bj{\o}rlykke and Halvorsen, P{\aa}l and Kielland-Gyrud, Halvor and Stensland, H{\aa}kon Kvale and Griwodz, Carsten},
  journal={ACM Transactions on Computer Systems (TOCS)},
  volume={38},
  number={1-2},
  pages={1--78},
  year={2021},
  publisher={ACM New York, NY, USA}
}

@article{liu2022improving,
  title={Improving fairness for SSD devices through DRAM over-provisioning cache management},
  author={Liu, Renping and Tan, Zhenhua and Long, Linbo and Wu, Yu and Tan, Yujuan and Liu, Duo},
  journal={IEEE Transactions on Parallel and Distributed Systems},
  volume={33},
  number={10},
  pages={2444--2454},
  year={2022},
  publisher={IEEE}
}

@inproceedings{hu2015pass,
  title={PASS: A proactive and adaptive SSD buffer scheme for data-intensive workloads},
  author={Hu, Yang and Jiang, Hong and Feng, Dan and Luo, Hao and Tian, Lei},
  booktitle={2015 IEEE International Conference on Networking, Architecture and Storage (NAS)},
  pages={54--63},
  year={2015},
  organization={IEEE}
}

\end{document}